\documentclass[12pt,preprint]{aastex}
\slugcomment{To appear in the Astronomical Journal}

\lefthead{Skinner et al.}
\righthead{Wolf-Rayet Stars}

\begin{document}

\title{X-ray Emission from  Nitrogen-Type  
            Wolf-Rayet Stars}

\author{Stephen L. Skinner}
\affil{Center for Astrophysics and Space Astronomy (CASA), 
       Univ. of Colorado, 
       Boulder, CO 80309-0389; email: Stephen.Skinner@colorado.edu }

\author{Svetozar A. Zhekov\altaffilmark{1}}
\affil{JILA, Univ. of Colorado,
       Boulder, CO 80309-0440 }

\author{Manuel G\"{u}del}
\affil{Institute of Astronomy, ETH Z\"{u}rich, 
       Wolfgang-Pauli-Str. 27,
       8093 Z\"{u}rich,
       Switzerland}

\author{Werner Schmutz}
\affil{Physikalisch-Meteorologisches Observatorium Davos and 
       World Radiation Center (PMOD/WRC),
       Dorfstrasse 33, 
       CH-7260 Davos Dorf, Switzerland}
\and

\author{Kimberly R. Sokal}
\affil{Center for Astrophysics and Space Astronomy (CASA), 
       Univ. of Colorado, 
       Boulder, CO 80309-0389 }

\altaffiltext{1}{On leave from 
       Space Research Institute, Sofia, Bulgaria}

% Notice that each of these authors has alternate affiliations, which
% are identified by the \altaffilmark after each name.  The actual alternate
% affiliation information is typeset in footnotes at the bottom of the
% first page, and the text itself is specified in \altaffiltext commands.
% There is a separate \altaffiltext for each alternate affiliation
% indicated above.

% The abstract environment prints out the receipt and acceptance dates
% if they are relevant for the journal style.  For the aasms style, they
% will print out as horizontal rules for the editorial staff to type
% on, so long as the author does not include \received and \accepted
% commands.  This should not be done, since \received and \accepted dates
% are not known to the author.
%
% Define symbol \ltsimeq
\newcommand{\ltsimeq}{\raisebox{-0.6ex}{$\,\stackrel{\raisebox{-.2ex}%
{$\textstyle<$}}{\sim}\,$}}
%
% Define symbol \gtsimeq
\newcommand{\gtsimeq}{\raisebox{-0.6ex}{$\,\stackrel{\raisebox{-.2ex}%
{$\textstyle>$}}{\sim}\,$}}
\begin{abstract}
We summarize  new  X-ray detections of four nitrogen-type 
Wolf-Rayet (WR) stars obtained in a limited survey aimed at
establishing the X-ray properties of WN stars across their
full range of spectral subtypes. None of the detected stars
is so far known to be a close binary. We report {\em Chandra} 
detections of WR 2 (WN2), WR 18 (WN4), and  WR 134 (WN6),
and an {\em XMM-Newton}  detection of WR79a (WN9ha).
These observations clearly demonstrate that both WNE and WNL 
stars are X-ray sources. We also discuss {\em Chandra} archive 
detections of the  WN6h stars WR 20b, WR 24, and WR 136 and
{\em ROSAT} non-detections of WR 16 (WN8h) and WR 78 (WN7h). 
The X-ray spectra of all WN detections show prominent emission 
lines and an admixture of cool (kT $<$ 1 keV) and 
hot (kT $>$ 2 keV) plasma. The hotter plasma is not 
predicted by radiative wind shock  models and other
as yet unidentified mechanisms are at work. Most stars
show X-ray absorption in excess of that expected from
visual extinction (A$_{\rm V}$),  likely due to their strong winds or 
cold circumstellar gas. Existing data suggest a falloff
in X-ray luminosity toward later WN7-9 subtypes, which
have higher L$_{bol}$ but slower, denser winds than
WN2-6 stars. This provides a clue that wind properties
may be a more crucial factor in determining emergent
X-ray emission levels than bolometric luminosity.
\end{abstract}
% The different journals have different requirements for keywords.  The
% keywords.apj file, found on aas.org in the pubs/aastex-misc directory, 
% contains a list of keywords used with the ApJ and Letters.  These are 
% usually assigned by the editor, but authors may include them in their 
% manuscripts if they wish. 


% That's it for the front matter.  On to the main body of the paper.
% We'll only put in tutorial remarks at the beginning of each section
% so you can see entire sections together.

% In the first two sections, you should notice the use of the LaTeX \cite
% command to identify citations.  The citations are tied to the
% reference list via symbolic KEYs.  We have chosen the first three
% characters of the first author's name plus the last two numeral of the
% year of publication.  The corresponding reference has a \bibitem
% command in the reference list below.
%
% Please see the AASTeX manual for a more complete discussion on how to make
% \cite-\bibitem work for you.   
\newpage

\section{Introduction}
X-ray observations of Wolf-Rayet (WR) stars provide an important means of
determining  physical conditions in their powerful metal-rich  winds and
in hot wind-blown bubbles which can surround  the star. 
X-ray observations yield information on the temperature, emission measure, 
and luminosity of hot plasma that can be used to test models of WR X-ray 
emission, which are largely based on the idea that the X-rays arise 
in shocked winds. Photoelectric absorption of low-energy X-rays can 
reveal the presence of gas along the line-of-sight toward the 
star that is difficult to detect by other means. High-resolution 
grating spectra can be obtained for the brightest WR X-ray sources 
and are capable of constraining the  location of the X-ray emittng 
plasma relative to the star, wind parameters, and wind element abundances.

It has been nearly three decades since WR stars were discovered
to be X-ray sources by the {\em Einstein Observatory} (Seward et al. 1979).
However,  the origin of the X-ray emission is still not totally
understood and more than one process may be involved. In the case of
single WR stars (and single OB stars), most theoretical work has focused
on  X-ray production in shocks that are formed in the wind via line-driven
instabilities (Lucy \& White 1980; Lucy 1982; Owocki, Castor, \&
Rybicki 1988; Baum et al. 1992; Feldmeier et al. 1997; Gayley \& Owocki 1995).
Such X-ray emission is expected to be soft (kT $\ltsimeq$ 1 keV) and
X-ray emission lines lines in optically thick winds are predicted to be 
asymmetric with blueshifted centroids (Owocki \& Cohen 2001). Alternate
models which attribute shock production to plasmoids moving through the 
wind have also recently been proposed (Waldron \& Cassinelli 2009).
The emission of WR binaries is potentially more complex than
single stars since  it may consist of intrinsic stellar wind-shock 
emission plus X-rays from a colliding wind shock between the two stars 
(Prilutskii \& Usov 1976; Luo et al. 1990; 
Stevens, Blondin, \& Pollock  1992; Usov 1992). The maximum X-ray 
temperature of an adiabatic  colliding  wind shock is proportional
to the square of the wind speed perpendicular to the shock front 
(Luo et al. 1990).  Values kT$_{max}$ $\gtsimeq$ 2 keV are expected for 
typical WR  wind speeds of v$_{\infty}$ $\sim$ 1000 - 5000 km s$^{-1}$.

Attempts to confirm X-ray wind-shock models observationally have
yielded mixed results. Some evidence supporting the colliding
wind picture in massive binaries  has been obtained from {\em Chandra} 
and {\em XMM-Newton} grating observations of X-ray bright systems
such as $\gamma^2$ Vel (WC8 $+$ O7.5; Skinner et al. 2001; 
Schild et al. 2004) and WR 140 (WC7 $+$ O4-5; Pollock et al. 2005).  
On the other hand, the radiative wind shock picture for single massive
stars has not yet found broad observational support. Perhaps the
best evidence for X-ray production in radiative wind shocks has 
come from grating observations of the O4f supergiant $\zeta$ Pup, 
which reveal wind-broadened emission lines 
(HWHM $\approx$ 600 - 1600 km s$^{-1}$) with blueshifted 
centroids (Cassinelli et al. 2001) and a
substantial emission measure contribution from low-temperature
plasma (Kahn et al. 2001). However, grating spectra of other 
OB stars show narrower unshifted symmetric emission lines that are more
difficult to reconcile with  radiative wind shock predictions
(e.g. Miller et al. 2002; Schulz et al. 2000; Cohen et al. 2008; 
Skinner et al. 2008). 

Compelling evidence for X-rays from radiative wind shocks in single WR
stars  is currently lacking.
But, very few sensitive pointed observations of single WR stars
exist and there are no single WR stars yet known that are bright
enough in X-rays to acquire good-quality  grating spectra
in reasonable exposure times. 
Even so, CCD observations without gratings are feasible and
moderate resolution CCD spectra capable of 
distinguishing between cool plasma that could arise in 
radiative wind shocks and hotter plasma due to other processes
such as colliding winds can be readily obtained. We have thus initiated
an X-ray  survey of single WR stars with  {\em Chandra} and 
{\em XMM-Newton} aimed at broadening the observational data 
base and determining their basic properties such as X-ray temperature
and luminosity. Our initial targets included both nitrogen-rich
WN stars and carbon-rich WC stars. They were selected from the 
VIIth Catalog of Galactic Wolf-Rayet stars (van der Hucht 2001,
herafter vdH01), giving preference to objects lying closest to the Sun with 
low A$_{\rm V}$ and no evidence for binarity. 
 
So far, we have observed four apparently single carbon-type
WC stars and, surprisingly, none was 
detected (Skinner et al. 2006; 2009). 
These are WR 5 (WC6), WR 57 (WC8), WR 90 (WC7), and WR 135 (WC8).
The most stringent 
upper limit is from a {\em Chandra} observation of WR 135 
for which we obtained log L$_{\rm X}$(0.5 - 7 keV)  $\leq$ 
29.82 ergs s$^{-1}$. This is $\approx$2 - 3 orders of 
magnitude below typical X-ray emission levels of WR binaries
and X-ray detected WN stars that are not known to be binaries.
A non-detection of the WC5 star WR 114 was also reported by 
Oskinova et al. (2003). Sensitive observations of a broader
sample of single WC stars are still needed, but the lack of
detections in these exploratory observations already suggests
they are either intrinsically faint X-ray emitters, or perhaps
X-ray quiet.

In contrast, high-confidence X-ray detections have been 
obtained for some apparently single WN stars (Skinner et 
al. 2002a,b; Ignace et al. 2003a; Oskinova  2005;
Naz\'{e} et al. 2008). These detections are confined primarily 
to spectral subtypes of WN6 and earlier. Very little is yet 
known about the X-ray properties of  WN7-9 stars. The 
X-ray luminosities of subtypes WN6 and earlier span a wide
range log L$_{\rm X}$ $\sim$ 31 - 33 ergs s$^{-1}$. 
It is remarkable that WN6 stars themselves have L$_{\rm X}$ 
values spanning nearly two orders of magnitude.
In those cases where good X-ray spectra of early WN subtypes
($\leq$WN6) exist, a multi-temperature plasma is invariably
present with a cool component at kT$_{1}$ $\ltsimeq$ 0.6 keV 
and a hotter component at kT$_{2}$ $\gtsimeq$ 2 keV.  The
cool plasma is compatible with radiative wind shock predictions
but the hot plasma is not.

The existence of X-ray emission in late WN7-9 stars has not
been thoroughly investigated. But, there are indications that
their X-ray luminosity levels are low. An  {\em XMM-Newton} 
observation of WR 40 (WN8; Gosset et al. 2005) yielded a 
non-detection with a conservative upper limit
log L$_{\rm X}$  $\leq$ 31.6 ergs s$^{-1}$. But, an 
archival {\em ROSAT} PSPC image (rp200150n00) reveals a marginal
detection of the WN9 star WR 79a  at a 12.2$'$ off-axis position,
which was our primary motivation for reobserving it at higher
sensitivity with {\em XMM-Newton}.

We report here on sensitive X-ray observations of 
four  WN stars of spectral subtypes WN2 - WN9, namely
WR 2 (WN2), WR 18 (WN 4), WR 134 (WN6), and WR 79 (WN9ha). 
None of these
is listed as either a binary or probable binary in the VIIth 
WR Catalog (vdH01). 
Periodic line-profile spectroscopic variability in WR 134
at a period P = 2.25 d has been reported  by Morel et al. (1999),
but no companion has yet been found and other factors such
as a rotationally-modulated wind may be responsible for the 
variability (St.-Louis et al. 2009).  All four WN stars were 
detected (Figs. 1 - 4).
We supplement these new detections with selected X-ray archival 
data for other WN stars, summarize their X-ray properties, and
discuss candidate  emission mechanisms. Table 1 summarizes the
general properties of the WN stars considered in this study.

\section{Observations}

\subsection{{\em Chandra}}
The {\em Chandra} observations are summarized in Table 2.
Exposures were obtained using the ACIS-S (Advanced CCD 
Imaging Spectrometer) imaging array in faint 
timed-event mode. A reduced 1/4 subarray size was
used for the brighter source WR 134 to minimize
any photon pileup. The WR star target was placed
at the nominal aimpoint on CCD S3 of the array.
The CCD pixel size is 0.492$''$.
Approximately 90\% of the encircled energy at 1.49 keV
lies within 2$''$ of the center pixel of an on-axis  point source. 
More information on {\em 
Chandra} and its instrumentation can be found in the {\em Chandra} Proposer's 
Observatory Guide (POG)\footnote {See http://asc.harvard.edu/proposer/POG}.

The Level 2 events file provided by the {\em Chandra} X-ray
Center (CXC) was analyzed using standard science
threads in CIAO 
version 4.1.2\footnote{Further information on 
{\em Chandra} Interactive
Analysis of Observations (CIAO) software can be found at
http://asc.harvard.edu/ciao.}.
The CIAO processing  used calibration
data from CALDB version 4.1.2.
Source detection was carried out using the 
the CIAO {\em wavdetect} tool, which correlates
the input image with ``Mexican Hat'' wavelet
functions of different scale sizes.
We ran {\em wavdetect} on full-resolution images with
a pixel size of 0.$''$492 using  events in the 0.3 - 8 keV
range to reduce the background. The {\em wavdetect}
threshold was set at $sigthresh$ = 1.5 $\times$ 10$^{-5}$ 
and scale sizes of 1, 2, 4, 8, and 16 were used.
The {\em wavdetect} tool provides source centroid
positions and net source counts (background subtracted)
inside the computed 3$\sigma$ source region (Table 3).

CIAO {\em specextract} was used to extract  source and background 
spectra for the WN stars along with source-specific
response matrix files (RMFs) and auxiliary response files (ARFs).
We used the 3$\sigma$ source ellipses from
{\em wavdetect} to define the extraction regions and background was
extracted from adjacent source-free regions. Background is negligible,
contributing $<$1 count  (0.3 - 8 keV) inside the source
extraction regions over the duration of each observation.
Spectral fitting was undertaken with the HEASOFT 
{\em Xanadu}\footnote{http://heasarc.gsfc.nasa.gov/docs/xanadu/xanadu.html.}
software package including XSPEC vers. 12.4.0.
X-ray light curves were extracted using the CIAO tool
{\em dmextract} and checks for source variabilility
were carried out on energy-filtered source event files using the
Kolmogorov-Smirnov (KS) test (Press et al. 1992) 
and the Bayesian-method CIAO tool {\em glvary} 
(Gregory \& Loredo 1992, 1996).

\subsection{{\em XMM-Newton}}

Table 2 summarizes the {\em XMM-Newton} observation of WR 79a.
Data were acquired with the European Photon
Imaging Camera (EPIC), which provides charge-coupled
device (CCD) 
imaging spectroscopy from the pn camera
(Str\"{u}der et al. 2001) and two nearly
identical MOS cameras (MOS1 and MOS2;
Turner et al. 2001). The observation was
obtained in full-window mode using the thick
optical blocking filter. The EPIC cameras
provide  energy coverage in the range
E $\approx$ 0.2 - 15 keV with energy 
resolution E/$\Delta$E $\approx$ 20 - 50.
The MOS cameras provide the best on-axis angular 
resolution with FWHM $\approx$ 4.3$''$ at
1.5 keV.

Data were reduced using the {\em XMM-Newton}
Science Analysis System (SAS vers. 7.1) using
the latest calibration data. Event files based on
pipeline processing carried out at the
{\em XMM-Newton } Science Operations Center were 
filtered to select good event patterns. Further
filtering based on event energies was done to
reduce background emission. No time filtering
was required. 

Spectra and light curves were extracted from  circular
regions  of radius  R$_{e}$ = 15$''$ centered
on WR79a, corresponding to $\approx$68\%
encircled energy at 1.5 keV. This approach reduces
the number of background counts while capturing
most of the source counts. Total source counts 
(Table 3) were measured in a larger circular 
region of radius r = 45$''$ corresponding to
$\approx$90\% encircled energy at 1.5 keV.
Background spectra and light curves were obtained
from circular source-free regions near
the source. The SAS tasks
{\em rmfgen} and {\em arfgen} were used to
generate source-specific 
RMFs and ARFs for spectral analysis. The data were
analyzed using the HEASOFT {\em Xanadu} 
software package

\subsection{Archival Data}

We have examined {\em Chandra} and {\em ROSAT} archive
data for selected WN stars which are not known to be binaries.
Our archive analysis is not exhaustive and was 
restricted to a few high-interest objects for which
either good CCD spectra or useful upper limits were
obtained.  Analysis of {\em Chandra} archive data was carried
out using the same procedures described above (Sec. 2.1).
{\em ROSAT} images were analyzed using the {\em Ximage}
image analysis tool, which is part of the 
HEASOFT {\em Xanadu} package. Results of the archive
analysis are summarized in Section 3.3 and in Tables 3 and 5.

\section{Results }

\subsection{General X-ray Properties}

Table 3 summarizes the X-ray properties of 
the four newly-detected WN stars and three
WN stars from the {\em Chandra} archive
(WR 20b, WR 24, and WR 136). There is very
good  agreement between the  X-ray 
positions and the optical positions of the WN stars
given in the {\em HST} Guide Star Catalog (GSC).
Radial offsets between the X-ray 
and optical positions are $\leq$0.3$''$ for  the
new detections, being  within the 1$\sigma$ 
range of {\em Chandra} 
\footnote {http://asc.harvard.edu/proposer/POG}
and {\em XMM}
\footnote{http://xmm2.esac.esa.int/docs/documents/CAL-TN-0018.pdf}
positional accuracy. Similar offsets were obtained for the
archive detections. There is thus no reason to doubt that the
X-ray sources are the optically-identified WR
stars. 

No significant X-ray variability was
detected in any of the targets, but  sampling in
the time domain spans less than $\approx$10 hours for 
the four new detections and $\ltsimeq$1 day for the archive sources.
Thus, variability on longer timescales exceeding one day
cannot yet be ruled out. 

The X-ray spectra of the four new detections are shown in 
Figure 5 and the archive spectra in Figure 6. All spectra 
show some absorption below $\approx$0.8 keV as well as 
prominent emission lines or line complexes indicative of 
thermal emission.  The most prominent lines
are the unresolved Mg XI He-like triplet at  1.33 - 1.35 keV
(maximum line power temperature  log T$_{max}$ = 6.8 K) and
the unresolved Si XIII triplet at 1.84 - 1.86 keV
(log T$_{max}$ = 7.0 K). The higher temperature S XV line
at 2.46 keV (log T$_{max}$ = 7.2 K) is also seen in all  
stars, but is very faint in WR 2. The high-temperature
line  from He-like Ar XVII (log T$_{max}$ = 7.3 K), is detected in
several stars and is quite strong in WR 20b. Faint emission
near 3.9 keV in the WN6h stars  WR 20b and WR 136 may
be due to He-like Ca XIX (log T$_{max}$ = 7.5 K) but we
classify this line only as a possible detection.
The Fe K$\alpha$ complex
(Fe XXV; 6.67 keV log T$_{max}$ = 7.6 K) is detected in
WR 20b, WR 134, and WR 136. This line is a signature of
very  hot plasma, which is undoubtedly present in
these latter three WN6h stars.

\subsection{X-ray Spectral Fits}

We have attempted to fit the spectra with various emission 
models, including the XSPEC variable-abundance optically thin 
plasma model $vapec$ (Smith et al. 2001) and the constant
temperature plane-parallel shock model
$vpshock$ (Borkowski et al. 2001, and references therein).
All models included a photoelectric absorption component
(XSPEC model $wabs$) which was used to determine the 
equivalent neutral H column density N$_{\rm H}$.

Element  abundances are expected to deviate strongly
from solar composition in WN stars as the result of advanced 
nucleosynthesis. H is depleted along with  C and O, while He and
N are enriched (van der Hucht et al. 1986; hereafter VCW86).
Because of the moderate CCD spectral resolution of ACIS-S and
EPIC and limited total source counts, we are not able to place 
tight constraints on  element abundances using the X-ray spectra. 
We have thus adopted the generic WN abundances of VCW86  
(Table 4) as starting values in  spectral fits using the
$vapec$ and $vpshock$ models.
Selected elements were allowed to deviate from the VCW86
values to improve the fits, as detailed in Tables 4 and 5.
The total X-ray absorption likely includes constributions
from solar abundance material in the interstellar medium
and stellar wind absorption from material enriched with
heavy elements. The CCD spectra do not provide sufficient
information to disentangle solar and nonsolar absorption,
and we have thus modeled X-ray absorption using 
solar abundances (XSPEC $wabs$; Anders \& Ebihara 1982).

Our main spectral analysis results can be summarized as follows:
(i) single-temperature optically-thin plasma models (1T $vapec$) 
do not provide acceptable fits, even if abundances are allowed
to vary,  (ii) two-temperature models (2T $vapec$) are acceptable
but the goodness-of-fit statistic ($\chi^2$) is sensitive to abundances; 
(iii) all stars show both a cool (kT$_{1}$ $<$ 1 keV) and
hot  (kT $>$ 2 keV) plasma component, 
(iv) most stars show X-ray absorption in excess of that expected
from their optical extinction A$_{\rm V}$, (v) fits using isothermal 
plane-parallel shock models are not as good as 
2T $vapec$ (WR 20b excepted).

\subsubsection{Optically Thin Plasma Models ($vapec$)}

In all cases, the 2T model requires a cool component 
at kT$_{1}$ $\approx$  0.3 - 0.6 keV  plus a hotter 
component at kT$_{2}$ $>$ 2 keV.
The best-fit value of kT$_{2}$ for WR 2 is 
sensitive to the S abundance, which is not well-constrained
by the data. We thus quote only a lower 90\% confidence bound
on kT$_{2}$ for WR 2. The respective normalization values ($norm$) for
each component show that most of the emission measure  in WR 2, WR 18,
and WR 79a is associated with the cooler component, while the 
hotter component clearly dominates in WR 134 and WR 136.

The best-fit  neutral-H absorption column density (N$_{\rm H}$) 
is a factor of $\approx$2 - 4 lower in WR 2 than in the other 
three stars. Using the conversion 
N$_{\rm H}$  $=$ 2.22 $\times$ 10$^{21}$ A$_{V}$ cm$^{-2}$
(Gorenstein 1975), it is apparent from Table 4 that the X-ray
derived N$_{\rm H}$ is consistent with that expected from
A$_{V}$ = 1.58 mag for WR 2 (Table 1). However, the N$_{\rm H}$
for WR 18, WR79a,  and WR 134 is a factor of $\sim$2 - 4 greater than
anticipated from A$_{V}$, implying the presence of X-ray absorbing 
gas (e.g. the wind) that does not contribute significantly to
the optical extinction. Excess absorption is also present in 
the archive detection WR 20b.

Some improvement in the 2T $vapec$ fits was obtained by allowing
elements with detectable line emission (N, Ne, Mg, Si, S, Fe) to deviate from
the generic WN values (Table 4 and 5 notes). In particular, our fits of 
several stars were improved by using  S abundances larger 
than the value given in VCW86 (which was based on cosmic mass
fraction).

\subsubsection{Plane-Parallel Shock Models ($vpshock$)}

Given the expectation that WR star X-ray emission may arise
in shocked winds, we attempted to fit the spectra of all
detections except WR 18 with $vpshock$ in order
to determine if a simplistic
constant-temperature shock model could reproduce the spectra.  
There are insufficient counts in WR 18 to establish a clear 
preference between thin plasma models and shock models. 
The $vpshock$ fits used data from the Atomic Plasma 
Emission Database (APED)
\footnote{http://cxc.harvard.edu/atomdb/sources\_aped.html}
 to calculate the spectrum (XSPEC option $neivers$ 2.0).

For the new detections, reasonably good $vpshock$ fits were 
obtained for both WR 2 and WR 134, but we do not find a 
clear preference for $vpshock$ over the 2T $vapec$ model 
in either case.
For WR 2, $vpshock$ gives shock temperatures in the 
range kT$_{shock}$ $\approx$ 1.5 - 3.5 keV, but in order
to obtain  acceptable fits the  abundances of elements
with line emission above 2 keV (most notably S and Ar)  must 
be increased to very high values. When abundances are fixed
at the generic WN values, $vpshock$ underestimates the 
hard flux above 2 keV. When abundances are allowed to
vary, the model attempts to correct for this flux deficit
by running up the S and Ar abundances while at the same
time keeping kT$_{shock}$ $\approx$ 2 keV. This is likely
a consequence of attempting to fit a multi-temperature
plasma with an overly simplistic constant-temperature shock model.
For WR 134, high shock temperatures  kT$_{shock}$  
$\geq$ 10 keV are required and even with  such high 
temperatures the $\chi^2$ fit statistic is not as good 
as obtained with 2T $vpaec$.

For the archive detections, the  $vpshock$ fits for WR 24
and WR 136 are reasonably good above 1 keV but the shock
model underestimates the observed flux of the softer emission 
below $\approx$1 keV. WR 20b provides an interesting 
exception because the $vpshock$ model actually provides 
a better fit to its spectrum than  2T $vapec$ (Table 5 notes). 
The inferred  shock temperature  kT$_{shock}$ = 3.9  keV
is quite high. However, the
soft emission below 1 keV that cannot be accurately
reproduced with $vpshock$ in the other stars is not
detected in WR 20b (Fig. 6). Thus, the good results obtained with
$vpshock$ for WR 20b may simply be a consequence of its higher
absorption.

\subsection{Comments on Specific Stars}

\noindent{\bf WR 2 (WN2)}:~  
WR 2 is a new  {\em Chandra} detection and is the earliest
WN subtype in our sample. The V = 12.73 mag object 
({\em HST} GSC J010522.98$+$602505.1) lying 13.9$''$  south 
of WR 2 was not detected by {\em Chandra}. WR 2 lies in a neutral
hydrogen void (Arnal et al. 1999). No significant spectroscopic
variability was found for WR 2 in the recent study of 
St.-Louis et al. (2009). However, unusual round-shaped
emission line profiles have been reported for WR 2 
that were interpreted as signaling rapid rotation 
near breakup (Hamann et al. 2006).  

As Table 3 shows, it is the softest
X-ray source of the detected WN stars as judged by its  
median photon  energy E$_{50}$ = 1.04 keV.
Spectral analysis  confirms that the X-ray
absorption toward WR 2 is less than the other three stars,
resulting in the detection of more soft photons
below 0.7 keV and a lower median photon energy.
Events were detected in WR 2 down to energies of
$\approx$0.3 keV as well  emission at higher energies 
up to $\approx$3 keV. 

The {\em Chandra} spectrum is interesting because it
shows an emission feature near 0.5 keV that
likely includes a contribution from the low-temperature
hydrogen-like  N VII Ly $\alpha$ line (log T$_{max}$ = 6.3 K). 
The intermediate temperature Ne X line at 1.02 keV
(log T$_{max}$ = 6.7 K) is also detected, as well as
faint emission from the higher-temperature Si XIII
and S XV lines. Thus, cool and hot plasma features are
present in the same star.
Two issues related to 2T $vapec$ fits are worthy of note.
This model is able to reproduce most of the 
broad emission feature near 0.5 keV with the 
N VII Ly$\alpha$ line (E$_{lab}$ = 500.3 eV).
But, there is some residual at slightly higher energies of
530 - 540 eV that  is not reproduced by this line. 
The O VII resonance line is not obviously detected 
and in any case it lies at a higher energy (574 eV) and 
cannot account for this residual. However, other ions with lab
energies slightly above that of N VII could be contributing to
the residual including N VI (532.6 eV), S XIV (538.9 eV), and
Ca XVI (546.8 eV). The 2T $vapec$ model also has difficulty
reproducing all of the flux in the faint feature near
2.46 keV (S XV) unless a very high S abundance is assumed.

\noindent {\bf WR 16 (WN8h)}:~This star is of interest because of
its later WN8h subtype and moderate extinction A$_{\rm V}$ =
1.8 mag (vdH01).  WR 16 was not detected in  {\em ROSAT} PSPC 
image rp200715n00, which has an exposure livetime of 7,465 s.
Our analysis of the PSPC data gives a  3$\sigma$ upper limit 
on the total band count rate $\leq$1.8 c ksec$^{-1}$.
The PSPC is sensitive at lower energies of $\approx$0.15 -
2.5 keV. This equates to a conservative 3$\sigma$ upper limit
log L$_{\rm X}$(0.3 - 8 keV) $\leq$ 31.7 ergs s$^{-1}$,
assuming a distance  d = 2.37 kpc (vdH01).
The conversion from PSPC rate to unabsorbed flux and L$_{\rm X}$ was
carried out using the Portable Interactive Multi-Mission Simulator 
(PIMMS)\footnote{http://heasarc.nasa.gov/docs/software/tools/pimms.html}
assuming N$_{\rm H}$ = 4 $\times$ 10$^{21}$ cm$^{-2}$
corresponding to   A$_{\rm V}$ = 1.8 mag (Gorenstein 1975)
and a Raymond-Smith thermal plasma model with emission measure
equally distributed between a cool component at kT$_{1}$ = 0.6 keV
and a hotter component at kT$_{2}$ = 3.0 keV.

\noindent{\bf WR 18 (WN4)}:~
This WN4 star is associated with the asymmetric
nebula NGC 3199, which has been modeled as a
wind-blown bubble by Dyson \& Ghanbari (1989).
Cold circumstellar molecular
gas has been reported by Marston (2001, 2003).
It is a new {\em Chandra} detection and its spectrum shows a
prominent Si XIII line and the hotter S XV line.

\noindent{\bf WR 20b (WN6h)}:~
This star was detected as a bright
source in a previous {\em Chandra} observation of 
the massive stellar cluster
Westerlund 2 (ObsId 6411; Naz\'{e} et al. 2008). Its X-ray emission
is heavily-absorbed and exceptionally hard with a median  photon energy 
E$_{50}$ = 3.04 keV.  
We were unable to obtain acceptable fits with a 1T $vapec$ model
but a 2T $vapec$ model is satisfactory and requires a
hot plasma component at kT$_{2}$ $\approx$ 4 - 5 keV.
The isothermal plane-parallel shock model 
$vpshock$ provides a slightly
better fit with an inferred shock temperature
kT$_{shock}$ = 3.9 keV. The X-ray absorption is 
greater than for any other WN star in our sample.
The absorption determined from both  $vapec$ and
$vpshock$ models yields a ratio of X-ray to optically-determined
absorption N$_{\rm H}$/N$_{\rm H,vis}$ $\approx$ 3.
The ACIS-I spectrum shows high-temperature
emission lines including Ar XVII (3.13 keV) and Fe XXV (6.67 keV).
The distance to WR 20b is quite uncertain.
Values range from d = 2.27 kpc (vdH01)
up to 8 kpc (Naz\'{e} et al. 2008). Even at the low end of
this range, its X-ray luminosity is very high for a WN
star (log L$_{\rm X}$ $>$ 33.5 ergs s$^{-1}$). Its high L$_{\rm X}$ 
and hard spectrum are reminiscent of colliding wind binaries,
suggesting that this object may not be a single star (Sec. 4.6).

\noindent {\bf WR 24 (WN6ha)}:~This star was detected as a 
bright off-axis X-ray source in an archival {\em Chandra}  
guaranteed time observation (GTO) of the Carina region (GTO ObsId 9482).  
Its ACIS-I spectrum shows a remarkably strong S XV 
emission line at 2.46 keV, as well as the high-temperature
Ar XVII (3.13 keV) line and  possible faint Fe line
emission in the 6.4 - 6.7 keV range.  
A 2T $vapec$ model
provides a satisfactory fit with a hot plasma component
at kT$_{2}$ $\approx$ 4 keV. The absorption
N$_{\rm H}$ = 0.76 [0.44 - 1.05] $\times$ 10$^{22}$ cm$^{-2}$
is several times larger than would be expected from 
A$_{\rm V}$ = 0.56 mag (vdH01). At an assumed distance
d = 3.24 kpc (vdH01), a high  unabsorbed X-ray luminosity
log L$_{\rm X}$ = 33.0 ergs s$^{-1}$ is inferred, only
slightly less than for WR 20b. Because of its high
L$_{\rm X}$ and high L$_{bol}$, WR 24 is another
candidate unresolved binary system (Sec. 4.6).

\noindent {\bf WR 78 (WN7h)}:~This WNL star is similar to WR 16,
having moderate extinction A$_{\rm V}$ = 1.5 mag and 
is relatively nearby at d = 1.99 kpc (vdH01). A previous
tabulation of WN star X-ray observations by Oskinova  (2005)
listed WR 78 as a {\em ROSAT} PSPC detection. Our analysis of
archival  {\em ROSAT} PSPC images rp200716n00  does not confirm 
this. We report a non-detection in the total band and soft band
images. Weak signal may be present in the hard band PSPC
image, but the signal-to-noise ratio S/N $\approx$ 2 is not 
sufficient to classify it as a detection. We obtain a 
3$\sigma$ upper limit on the total band PSPC count rate
$\leq$2.97 c ksec$^{-1}$ based on an exposure livetime of
10,121 s. The corresponding 3$\sigma$ luminosity upper limit from PIMMS is
log L$_{\rm X}$(0.3 - 8 keV) $\leq$ 31.7 ergs s$^{-1}$ (at d = 1.99 kpc).
This assumes N$_{\rm H}$ = 3.3 $\times$ 10$^{21}$ cm$^{-2}$
corresponding to   A$_{\rm V}$ = 1.5 mag (Gorenstein 1975)
and a 2T Raymond-Smith thermal plasma model as for WR 16 above.

\noindent {\bf WR 79a (WN9ha)}:~This is a new and important
{\em XMM-Newton} detection, being the latest WN subtype
in our sample and the first unambiguous X-ray detection of 
a WNL star in the WN7-9 range. There is a fainter
near-IR source (2MASS J165458.09$-$410903.6) at an offset
of 4.$''$8 from WR 79a (see also Mason et al. 1998), but
the X-ray positional accuracy is capable of distinguishing
between this object and WR 79a. Radio continuum emission
has been detected (Bieging, Abbott, \& Churchwell 1989)
and follow-up observations suggest a nonthermal radio
component may be present (Cappa, Goss, \& van der Hucht 2004).
WR 79a (= HD 152408) was formerly classified as O8:Iafpe
but was subsequently reclassified as WN9ha based on the 
analysis of its optical, UV, and IR spectra by
Crowther \& Bohannan (1997). 
Its EPIC spectrum reveals a strong Si XIII line and 
higher temperature S XV and Ar XVII lines.
Since this star resembles
an O8 supergiant, we have considered solar abundance
spectral fits as well as the fit based on  WN
abundances given in Table 4. The solar abundance fit
gives comparable temperatures to the WN abundance fit
but yields a slightly higher N$_{\rm H}$ and 
L$_{\rm X}$ (Table 4 Notes). Regardless of the reference
abundances used, the L$_{\rm X}$ of WR 79a is at the low
end of the range found here for WN stars.

\noindent {\bf WR 134 (WN6h)}:~
This is a new {\em Chandra} detection and a luminous
X-ray source with a very hot component. It is 
surrounded by a ring nebula (Chu et al. 1983;
Miller \& Chu 1993). Because of its unusual 2.25 d 
periodic spectroscopic variability,
it has been the subject of numerous previous studies
(e.g. Howarth \& Schmutz 1992; Schulte-Ladbeck et al. 1992;
Morel et al. 1999; St.-Louis et al. 2009). 
The possibility that the 2.25 d  periodicity might
be due to a compact companion has been debated 
but there are difficulties with this interpretation
(Morel et al. 1999). We discuss the compact companion
hypothesis further in Section 4.4.

\noindent {\bf WR 136 (WN6h)}:~WR 136 is surrounded by the 
ring nebula NGC 6888 (Gruendl et al. 2000) which is a 
source of diffuse X-ray emission (Bochkarev 1988;
Gruendl, Guerrero, \& Chu 2003; Chu, Gruendl, \& Guerrero 2006). 
The WR star was detected
at  $\approx$8$'$ off-axis in a deep 93 ksec {\em Chandra} 
exposure of  NGC 6888 (ObsId 3763). The ACIS-S spectrum is faint, but
quite remarkable. The spectrum  is dominated almost entirely by emission 
lines,  including the high-temperature S XV line (2.46 keV), 
possible faint emission from Ca XIX (3.9 keV), and a 
feature near 6.67 keV that is likely Fe XXV. Attempts to fit the
spectrum with a 2T  $vapec$ model using a single absorption 
component were unstable and resulted in high runaway 
temperatures for the hot
component. A modified 2T  $vapec$ model that allows for
different absorptions toward the cool and hot component is
stable and provides a good fit (Table 5). The absorption
derived for the cool component is consistent with that
expected for previous extinction estimates 
A$_{\rm V}$ = 1.73 $\pm$ 0.32 mag (vdH01). However, the 
best-fit absorption of the hot component is a factor of 
$\approx$8 larger. The physical picture that emerges from
such a model is that the cool X-ray emission originates 
far from the star and incurs little or no wind absorption.
In contrast, the  hot component originates deep within the 
wind and much closer to the star and is subject to strong
attenuation by the wind. The inferred X-ray luminosity is 
sensitive to the temperature of the cool component but most fits
yield values of  log L$_{\rm X}$(0.3 - 8 keV) $\approx$ 31.5
ergs s$^{-1}$, assuming the {\em Hipparcos} distance 
d = 1.64 kpc. Thus, WR 136 seems to be of lower
X-ray luminosity than other WN6 stars in the sample.

\section{Discussion}

The new observational results discussed above provide the
most detailed picture to date of the X-ray emission of
apparently single WN stars. We comment below on specific
aspects of their X-ray emission and possible emission 
processes.

\subsection{X-ray Absorption}

The visual extinction A$_{\rm V}$ for the WN stars in our sample
is dominated by interstellar material along the 
line-of-sight. There is insufficient dust in the winds of
WN stars to significantly affect  A$_{\rm V}$, in contrast
to some WC stars whose dusty winds can absorb and re-radiate
stellar light (van der Hucht, Williams, \& Th\'{e} 1987).
The A$_{\rm V}$ values in Table 1
can be used to calculate an equivalent neutral H absorption
column density along the line-of-sight using the conversion
N$_{\rm H,vis}$ = 2.22 $\times$ 10$^{21}$A$_{\rm V}$ cm$^{-2}$
(Gorenstein 1975). Assuming that the winds of WN stars have a 
negligible effect on  A$_{\rm V}$, the resulting value of
N$_{\rm H,vis}$ is a measure of interstellar absorption. 
In Tables 4 and 5 we give the ratio of
the X-ray to optically-derived absorption column densities
N$_{\rm H}$/N$_{\rm H,vis}$. This ratio
is greater than unity for all WN stars except WR 2.
That is, the X-ray absorption is greater than expected based on
visual extinction estimates. The largest absorption excesses 
are found for WR 20b and WR 134. For the double-absorption 
model of WR 136, there is no significant absorption excess
for the cool component but a large excess is found for the 
heavily-absorbed hot component.

Most of the  excess X-ray absorption likely  occurs in
the metal-rich  WR wind, but in some cases cold 
circumstellar molecular gas may also contribute. It may
not be a coincidence that WR 2 has little or no
excess absorption given that it has the highest effective
temperature of the four newly-detected stars (T$_{*}$ = 141 kK,
Hamann et al. 2006) and the lowest wind density parameter
$\dot{M}$/v$_{\infty}$. Higher temperatures (along with 
higher ionization) plus lower densities would increase
wind transparency to X-rays.

If the X-rays are formed in the wind then some 
wind absorption is expected, especially for 
lower energy photons. For an ideal spherical
wind, the radius of  optical 
depth unity R$_{\tau=1}$ beyond which X-ray photons
can escape (the ``exospheric approximation'') 
depends on several factors 
including photon energy, mass-loss parameters,
photoelectric absorption cross-sections, and 
wind abundances (e.g. Owocki \& Cohen 1999).
In Figure 7, we plot R$_{\tau=1}$ as a function of
photon energy for the four new detections. We have assumed
an ideal spherical, homogeneous, constant-velocity 
wind with the mass-loss 
parameters as given in Table 1, canonical WN
abundances (VCWH86), and the photoelectric 
absorption cross-sections of 
Baluci\'{n}ska-Church \& McCammon (1992, hereafter BM92).
It is apparent that under the above assumptions, soft
photons with E $<$ 1 keV emerge at large distances from
the star R$_{\tau=1}$ $\gtsimeq$ 1000 R$_{\odot}$.
Taking WR 134 as a specific example, the softest
photons visible in its spectrum (Fig. 5) at 
E $\approx$ 600 eV have R$_{\tau=1}$(600 eV) = 
9262 R$_{\odot}$ $\approx$ 1750  R$_{*}$, 
assuming R$_{*}$ = 5.29 R$_{\odot}$
(Hamann et al. 2006).

The above estimates, which are based on the assumption
that the wind is homogeneous and spherically symmetric,
suggest that the softest X-ray photons detected in the 
WN star spectra emerge at large distances of thousands 
of stellar radii or more from the star. These are only
rough estimates because wind abundances and 
mass-loss parameters are poorly-known. Furthermore,  
there is accumulating evidence that the winds of
WR stars are clumped rather than homogeneous
(Moffat et al. 1988; Robert 1994; L\'{e}pine et al. 2000),
in which case the X-rays could escape from smaller
radii closer to the star.
Even if the wind is the dominant absorber,
other factors such as cold dense circumstellar
gas could also contribute. Such gas has been detected
around  WR 18 and may be a vestige
of mass-loss in previous evolutionary 
phases (Marston 2001; 2003).

\subsection{Wind Shocks}

The radiative wind shock picture cannot explain the high-temperature
plasma seen in the X-ray spectra of WN stars, but the temperature of the
cooler plasma kT$_{1}$ $\approx$ 0.3 - 0.6 keV is compatible with radiative 
wind shocks.  Of the four stars observed, WR 2 shows the coolest 
emission. Assuming its cool plasma originates in wind shocks, 
we equate the best-fit temperature of the cool
component with the shock temperature, that is
kT$_{1}$ = 0.32 [0.24 - 0.45] keV = kT$_{s}$. 
For an adiabatic shock the maximum shock temperature
is  kT$_{s}$ = (3/16)\={m}v$_{s}^2$, where
v$_{s}$ is the shock velocity perpendicular to the 
shock front (Luo et al. 1990).
For a helium-rich WN wind \={m} = (4/3)m$_{p}$ 
where m$_{p}$ is the proton mass. An equivalent
form of this relation that is easier to apply is
kT$_{s}$ = 2.61[v$_{s}$/1000 km s$^{-1}$]$^2$~keV.

The inferred shock speed for WR 2 is then
v$_{s}$ = 350 [304 - 416, 90\% confidence range] km s$^{-1}$.
This value is plausible, being only about 
one-fifth the terminal wind speed (Table 1) and
within  the range of shock velocity jumps
found in numerical simulations of radiatively driven
winds in O-type stars (Owocki et al. 1988;
Feldmeier et al. 1997). The slightly higher
values kT$_{1}$ $\approx$ 0.6 keV for the other
three stars give shock speeds v$_{s}$ $\approx$
480 km s$^{-1}$. 

The unabsorbed flux of the cool
X-ray component in WR 2 (Table 4) gives a cool-component 
luminosity log L$_{\rm X,1}$ = 32.47 (ergs s$^{-1}$)
and log [L$_{\rm X,1}$/L$_{wind}$] = $-$4.94. So
there is sufficient wind kinetic energy to power
the soft X-ray emission even at a rather low 
conversion efficiency from kinetic to radiative energy.
But, numerical models show that forward shocks 
decay with radius (Feldmeier et al. 1997) and it
remains to be demonstrated by detailed simulations 
that X-ray emitting wind shocks can persist as coherent
structures in WR winds out to large distances of hundreds
to thousands of stellar radii (Sec. 4.1).

We now consider another star, WR 20b, which is an interesting
case because of the good fit obtained with an isothermal 
plane-parallel shock model. The best-fit shock temperature 
kT$_{s}$ = 3.86 [2.97 - 6.53] keV is too high
to explain by radiative wind shocks, but is plausible
for colliding wind shocks. If the wind of WR 20b is 
shocking onto an unseen companion, then the inferred
shock velocity  perpendicular to the interface is
v$_{s}$ = 1216 [1067 - 1582] km s$^{-1}$. This estimate is for
an adiabatic shock and assumes a pure He wind for the 
WN star. The above value is reasonable, being comparable to
the terminal wind speeds of WN6 stars (Crowther 2007;
Hamann et al. 2006).

\subsection{Magnetically-Confined Winds}

There are at present no detections of magnetic fields in 
WR stars to our knowledge.  But, if magnetic  fields of sufficient 
strength exist then they could confine the ionized wind into two
oppositely directed streams which collide near the magnetic
equator and form a magnetically-confined wind shock (MCWS).
This mechanism is of interest because it is capable of 
producing hot X-ray plasma over a range of temperatures
up to  several keV  (Babel \& Montmerle 1997) and could 
potentially explain the hotter plasma seen in the 
WN star spectra.

A rough estimate of the field strength needed to confine
the wind can be obtained using the confinement parameter
$\eta$ = B$_{eq}^2$R$_{*}^2$/$\dot{M}$v$_{\infty}$
where B$_{eq}$ is the field strength at the magnetic 
equator (ud-Doula \& Owocki 2002). For $\eta$ $\approx$ 1
the wind is  marginally-confined by the B-field,
for  $\eta$ $<<$ 1 the wind overpowers the 
field and is not confined, and for  $\eta$ $>>$ 1 the
wind is strongly-confined. Assuming  $\eta$ $\approx$ 1
(marginal confinement) and adopting the  mass-loss parameters 
and stellar radii in Table 1,  one obtains magnetic
field strengths in the range 
B$_{eq}$ $\approx$ 3.5 kG (WR 134) up
to 17 kG (WR 2). These fields are for marginal  
confinement and larger fields would be needed for
strong confinement. Without concrete evidence 
for strong B fields in WR stars, this mechanism
must be considered speculative.

\subsection{Close Binary Companions }

Our four new detections and archive sources were selected 
on the basis that they are so far not known to be 
close (spectroscopic) binary systems. Thus, any attempt to 
explain their  X-ray emission by invoking an unseen companion 
is not strongly motivated on observational grounds.
However, it is worth keeping in  mind that close companions
at sub-arcsecond spacings are  difficult to detect, 
especially when in low-inclination orbits.

Of the WN stars in our sample, WR 134 has been regarded
as a possible close binary system because of its known
2.25 day  periodic line profile changes.
This issue has been previously discussed by 
Morel et al. (1999). However, binarity has not yet
been demonstrated and a  rotationally-modulated
wind is an alternative explanation for the
variability (St.-Louis et al. 2009).
A close companion, if present, could potentially explain
the hard X-ray emission in WR 134. Hard X-rays could be
produced by the WR wind accreting onto a compact object
or by the wind shocking onto the surface of a normal
(nondegenerate) stellar companion.  But this interpretation
encounters some difficulties. 

The most compelling example known so far of a WR star
accreting onto a compact object is Cyg X-3. It has
a high X-ray luminosity L$_{\rm X}$ $\sim$ 10$^{38}$ 
ergs s$^{-1}$ and the companion is thought to be either
a black hole or a neutron star  
(van den Heuvel \& De Loore 1973; Schmutz, Geballe, 
\& Schild 1996; Lommen et al. 2005). If the putative
companion of WR 134 were a neutron star, then  the 
predicted X-ray luminosity from wind accretion is
log L$_{\rm X,acc}$ $\sim$ 10$^{37}$ ergs s$^{-1}$
(Morel et al. 1999). This exceeds the observed value 
for WR 134 (Table 4) by more  than four orders of
magnitude. This interpretation is thus untenable 
unless some  mechanism is operating to inhibit
wind accretion (see Davidson \& Ostriker 1973 or
Lipunov 1982 for details on such mechanisms).
A similar luminosity mismatch has also been noted 
for the WN star WR 6 (EZ CMa) by Skinner et al. (2002b).
It also undergoes periodic optical variability 
(P = 3.76 d; Firmani et al. 1980) of unknown origin and a neutron star 
companion has been suggested as a possible explanation.
In the case of
WR 134, there are other obstacles to overcome if
the putative companion is a neutron star of typical
mass M$_{ns}$ $\approx$ 1.4 M$_{\odot}$. 
If P = 2.25 d is interpreted as an orbital period, 
then Kepler's third law gives a separation $a$ $\approx$ 20 R$_{\odot}$
$\approx$ 4 R$_{*}$. Here we have adopted  
M$_{WR}$ = 19 M$_{\odot}$ and   radius 
R$_{*}$ = 5.3 R$_{\odot}$ for WR 134
(Hamann et al. 2006). Unless the wind is inhomogenous,
X-rays at energies E $\approx$ 2 - 3 keV (obviously present 
in the spectrum) could not escape from  such small radii.
Specifically,  if we use the mass-loss parameters for WR 134 
in Table 1 and 
canonical WN abundances (VCW86), then R$_{\tau=1}$(2 keV)
$\approx$ 89  R$_{*}$. The slightly different mass-loss
parameters for WN6 stars in Crowther (2007) give
R$_{\tau=1}$(2 keV) $\approx$ 36 R$_{*}$. 

Given the luminosity mismatch  obtained in the neutron star
companion model, it seems more likely that any unseen 
companion is a normal (nondegenerate) star. It would
have been difficult for a massive companion 
(M$_{comp}$ $\approx$ M$_{WR}$) to have escaped detection,
but a lower mass star (M$_{comp}$ $<<$ M$_{WR}$)
would be more difficult to detect. The possibility of
a low-mass companion was previously considered in 
our analysis of the similar WN-type star WR 6
(Skinner et al. 2002b). Likewise, for WR 134 it can 
be shown that the hard-component X-ray luminosity  
could be accounted for by the WR wind impacting the 
surface of a lower-mass star of subsolar radius
(Usov 1992). Furthermore, the plasma temperature of the
hot component of WR 134  is compatible with 
maximum values expected from a shocked wind at or
near  the terminal wind speed of WR 134. Specifically,
v$_{\infty}$  $\approx$ 1700 - 1960 km s$^{-1}$ 
(St.-Louis et al. 2009) gives
kT$_{s}$ $\approx$ 7.5 - 10. keV. However, 
such a low-mass companion would be located very
close to WR 134 if P = 2.25 d is the orbital 
period. Separations of $<$5 R$_{*}$(WR) are
inferred from Kepler's third law, so again the 
escape of X-rays through the wind is  
potentially problematic. This problem is  mitigated
if the wind is clumped or if an unseen companion
is present at a larger separation than inferred 
above, resulting in a lower WR wind density at or
near the companion surface.

\subsection{Emerging Trends in WN Star X-ray Emission}

The sample of single WN stars for which X-ray parameters
have been reliably determined  from good-quality CCD spectra 
is not yet large enough to establish any clear correlations
between stellar and X-ray properties. Other factors such as
uncertain distances, bolometric luminosities,  and mass-loss 
parameters hinder correlation studies and can mask dependencies.  

However, X-ray observations now span  the  full range 
of  spectral subtypes from WN2  to WN9 and a few notable 
trends are  beginning to emerge.
As Figure 8 shows, there is an apparent falloff in
L$_{\rm X}$ toward later WN7-9 subtypes. The WN9ha
star WR 79a has the lowest  L$_{\rm X}$ of 
any detected single WN star so far. And, the WN8h star
WR 40 was undetected in a previous 
{\em XMM-Newton} observation at an upper limit 
log L$_{\rm X}$ $<$ 31.6 ergs s$^{-1}$ (Gosset et al. 2005). 
This upper limit is a factor of 3.5 below the L$_{\rm X}$
determined from the detection of WR 79a. 
Figure 8 also includes upper limits for 
WR 16 (WN8h) and WR 78 (WN7h) based on short-exposure
{\em ROSAT} PSPC archive images. Since {\em ROSAT}
was not sensitive to hotter plasma above $\approx$2.5 keV,
which is likely present, the PSPC upper limits 
should be considered tentative
until observations spanning a broader X-ray bandpass
can be obtained. 

The apparent decrease in L$_{\rm X}$ for 
WN7-9 subtypes is  surprising because
it is in the opposite sense of L$_{bol}$,
which is larger for WN7-9 stars than
WN2-6 stars  (Table 1; Hamann et al. 2006; Crowther 2007).
We plot L$_{\rm X}$ versus L$_{bol}$ in Figure 9.
This figure clearly illustrates that even though
the WN and WC stars observed to date have similar
L$_{bol}$, the WC stars (all undetected) are at least 
1 - 2 orders of magnitude less luminous in X-rays. 
Considering only the detected WN stars, one might be
tempted to argue for a weak increase in  L$_{\rm X}$
with L$_{bol}$. But, the absence of
X-ray detections for three WNL stars
with high L$_{bol}$ (WR 16, 40, 78) is clearly
at odds with this  picture. This result is
contrary to the trend seen in OB stars,
for which L$_{\rm X}$ generally increases with
L$_{bol}$, albeit with large scatter 
(Bergh\"{o}ffer et al. 1997). Thus, some 
other stellar parameters besides L$_{bol}$
are critical for determining  X-ray luminosity
levels in these WN stars, and wind properties 
may play a key  role.

The classical theory of hot star winds in the 
exospheric approximation predicts that 
L$_{\rm X}$ $\propto$ ($\dot{M}$/v$_{\infty}$)$^{1+s}$
where s = 1 for optically thin winds and 
$-1$ $<$ $s$ $<$ 0 for optically 
thick winds (Owocki \& Cohen 1999). The
existing WN star data do not show any clear
increase in  L$_{\rm X}$ with $\dot{M}$/v$_{\infty}$.
On the contrary, WR 79a is the least luminous 
X-ray source amongst the new detections, but has the
largest $\dot{M}$/v$_{\infty}$.

From the standpoint of observations, an increase
in L$_{\rm X}$ with wind kinetic energy would seem
to be more promising. In Figure 10, we plot  L$_{\rm X}$
versus L$_{wind}$ = (1/2)$\dot{M}$v$_{\infty}^2$
for several WN stars for which good X-ray
CCD spectra are available. 
Despite uncertainties in $\dot{M}$,
v$_{\infty}$, and distances, there is a noticeable 
increase in  L$_{\rm X}$ toward higher
values of L$_{wind}$ for the  stars 
considered in this study.

A notable outlier is  WR 136, which  is underluminous 
in X-rays by at least  a factor of $\sim$5 
compared to the other WN6 stars (Fig. 8)
and also has a low L$_{\rm X}$ for its wind
luminosity (Fig. 10). We have assumed 
d = 1.64 kpc   for WR 136 based on
its {\em Hipparcos} parallax. Other distance
estimates are less (e.g. d = 1.26 kpc, vdH01), 
so a distance underestimate is not likely to
be responsbile for its lower L$_{\rm X}$.
One way to correct for the low L$_{\rm X}$ is 
to postulate very cool undetected plasma (kT $<$ 0.2 keV)
that is masked by absorption. If such plasma 
is present, then values log L$_{\rm X}$ $>$
32.0 ergs s$^{-1}$ are possible.

Equally interesting is the non-detection of
WR 40 in X-rays. This is an important
result because it has a slow terminal wind
speed v$_{\infty}$ = 650 - 840 
km s$^{-1}$ (Hamann et al. 2006; Gosset et al. 2005)
and thus a high average wind density. This may be a clue
that X-ray emission ceases to be efficient
(or becomes totally absorbed by the wind) in WN stars 
with slow dense winds.

X-ray observations of a larger sample of single WN 
stars spanning a broader range of
mass-loss parameters  are needed to firmly
establish any dependencies between L$_{\rm X}$
and wind parameters. Observations of more
WNL stars with slow winds (v$_{\infty}$ $<$
1000 km s$^{-1}$) would be particularly
useful.

\subsection{Comments and Questions on Binarity}

Previous X-ray studies have shown that several WR $+$ OB or WR $+$ WR 
binaries have very high X-ray luminosities with typical values
log L$_{\rm X}$ $\gtsimeq$ 33.0 ergs s$^{-1}$. Early evidence
for this was found from {\em Einstein}  observations of WR stars. 
The median X-ray luminosity of seven WN binaries detected by 
{\em Einstein} was log L$_{\rm X}$(0.2 - 4 keV) = 32.92 ergs s$^{-1}$,
with a range of 32.76 - 33.53  ergs s$^{-1}$ (Pollock 1987).
More recent work has confirmed high X-ray luminosities in other
WN binary systems. For example, {\em XMM-Newton} observations of
the WN8 $+$ OB system WR 147 gave log L$_{\rm X}$(0.5 - 10 keV) = 
32.83 ergs s$^{-1}$ (Skinner et al. 2007). A {\em Chandra} observation
of the massive WN binary system WR 20a in Westerlund 2 yielded 
log L$_{\rm X}$ = 33.69 ergs s$^{-1}$ based on 1T spectral
fits (Naz\'{e} et al. 2008). Our analysis of the WR 20a 
{\em Chandra} data with 2T $vapec$ models gives even higher
values log L$_{\rm X}$(0.3 - 8 keV)  $>$ 34.0 ergs s$^{-1}$ 
for d $>$ 2.37 kpc. The closely-spaced WN5 $+$ O6 binary 
V444 Cyg (= WR 139) is also worthy of mention because of its
short $\approx$4.2 d orbital period. {\em ASCA} observations 
show that its X-ray luminosity varies with orbital phase,
reaching maximum values log L$_{\rm X}$(0.7 - 10 keV) $>$ 33.1
ergs s$^{-1}$ for an assumed distance of 1.7 kpc (Maeda et al. 1999).

Thus, there is no doubt that {\em some} WR binaries are X-ray luminous.
But, are {\em all} WR binaries X-ray luminous? We do not yet know the
answer to that question because it can be  very difficult to
distinguish a closely-spaced binary from a single WR star. It is very
likely that some WR stars currently classified as single objects are
in fact close unresolved binaries. Also, a high X-ray luminosity 
is not strictly required in a colliding wind WR binary system.
For an adiabatic colliding wind system the
luminosity scales roughly as 
L$_{\rm X}$ $\propto$ $\dot{M}^2$$v^{-3.2}$D$^{-1}$ where
$\dot{M}$ and $v$ are the mass-loss rate and wind-speed of 
the dominant component and D is the binary separation
(Luo et al. 1990; Stevens et al. 1992).
Binaries with lower mass-loss rates or wide separations 
are expected to have lower L$_{\rm X}$, all other factors
being equal.

Based on the above, which WN stars in our sample are good
binary candidates? WR 24 would appear to be a very good
candidate because of absorption lines in its optical 
spectrum (vdh01) and its
high  L$_{\rm X}$ and L$_{bol}$. In fact, its position in the 
(L$_{bol}$,L$_{\rm X}$) diagram is almost identical to that of
the known binary WR 147 (Fig. 9). WR 20b is also a good 
candidate because of its exceptionally high X-ray luminosity,
even when computed using the lowest current distance 
estimate of d = 2.37 kpc. Searches for close companions 
in these two stars are certainly  justified.

Apart from WR 20b and WR 24, there are no other stars in our sample
having high log L$_{\rm X}$ $\gtsimeq$ 33.0 ergs s$^{-1}$ that 
could signal binarity. A particularly interesting
case is WR 79a (WN9ha). This star would seem to be a good candidate
for binarity based on absorption lines in its spectrum and the 
possible presence of nonthermal radio emission (Cappa et al. 2004).
But its X-ray luminosity log L$_{\rm X}$ = 32.14 ergs s$^{-1}$
is at the low end of the detected WN stars. If WR 79a is indeed
a binary then its L$_{\rm X}$ is much less than  other known WN
binaries, or its distance has been underestimated. 

WR 134 is another intriguing example. Based on its known 2.25 day
spectroscopic variability it would seem to rank as a strong
binary candidate. As noted in Section 4.4, the high temperature 
of its hot X-ray component is indeed consistent with that expected 
if the wind of WR 134 were shocking onto a companion at terminal
speed.  But there are
questions as to whether the X-rays could escape if produced in
a closely-spaced colliding wind binary system, and the X-ray luminosity 
of WR 134 (log L$_{\rm X}$ = 32.66 ergs s$^{-1}$) is not as  high
as other known WN binaries. Thus, the existing X-ray data for
WR 134 do not provide an indisputable case for binarity. Further
observations of WR 134 aimed at determining whether its X-ray
emission is modulated at the 2.25 day optical period would be 
useful since colliding wind binaries such as $\gamma^2$ Velorum
and WR 140 do show orbital X-ray modulation
(Willis, Schild, \& Stevens 1995;  Zhekov \& Skinner 2000;
Pollock et al. 2005).

\subsection{WN Stars versus WC Stars: A Sharp Contrast in X-rays}

Our pilot survey has so far detected all  four WN stars
observed. In contrast, X-ray observations at similar sensitivity 
levels did not detect any of the four WC stars in our sample,
and a fifth went undetected in another {\em XMM-Newton} program
(Sec. 1). Why is it that most WN stars are detected as X-ray sources,
but WC stars are not?

An important clue to solving this puzzle may lie in the 
excess X-ray absorption seen in the spectra of most of the
WN stars (Sec. 4.1). Such excess absorption has been seen
in other WN stars such as WR 6 (WN4; Skinner et al. 2002b).
The wind is likely responsible for most of this excess. X-rays
produced within the wind must overcome the wind opacity
to escape, and the opacity increases toward lower
photon energies, all other factors being equal. Thus,
softer X-ray photons will be more heavily absorbed and
the softest photons detected will have emerged from the
wind at large distances from the star.

The wind opacity in WC stars is much higher than 
in WN stars due to metal-enrichment.
For a given set of mass-loss parameters, X-rays of a given
energy will have more difficulty escaping a WC wind than a
WN wind. As a specific example, we consider a generic
WR star with a typical mass-loss rate of  
$\dot{M}$ = 2 $\times$ 10$^{-5}$ M$_{\odot}$ yr$^{-1}$
and v$_{\infty}$ = 1800 km s$^{-1}$.
Adopting canonical WN and WC abundances (VCW86)
and BM92 cross-sections, the 
radius of optical depth unity for an X-ray photon
of energy  E =  1 keV is $\sim$13 times greater for a WC
star than for a WN star. Even at higher energies
E = 3 keV the R$_{\tau=1}$ ratio is still a 
factor of $\sim$11. Thus, unless X-rays of 
moderate energy E $\approx$ 1 - 3 keV are formed
very far out in the wind of a WC star (at thousands of
solar radii)  they will be absorbed.
This is perhaps the most plausible explanation for 
the absence of WC star X-ray detections at present.
But a larger sample of
WC stars needs to be observed in order to determine if
hard X-rays might escape through the wind in some cases.
The best candidates for such observations would be WC stars with 
higher T$_{eff}$ and lower wind densities (lower $\dot{M}$
and  higher v$_{\infty}$), which would conspire to reduce wind opacity.

\newpage
\section{Summary}

The most important results of this study are the following:

\begin{enumerate}

\item X-ray emission has been detected from four recently-observed
      WN stars ranging in spectral type from WN2 - WN9, none of which
      is so far known to be  a binary system. Archive data provide
      detections for an additional three WN6h stars. It is thus now 
      established that  both WNE and WNL stars are X-ray sources.

\item The spectra of all  stars are similar. All show an
      admixture of cool (kT$_{1}$ $<$ 1 keV) and hot  (kT $>$ 2 keV)
      X-ray plasma and prominent emission lines from ions such as
      Si XIII and S XV. X-ray luminosities are typically
      log L$_{\rm X}$ $\sim$ 32.5 ergs s$^{-1}$, but there is
      a rather large scatter of $\pm$1 dex around this value.
      Some of this scatter is likely due to uncertain distances
      for many WR stars.

\item Most of the detected WN stars show  X-ray absorption in excess of 
      that expected from published A$_{\rm V}$ estimates 
      (WR 2 being an exception). A plausible explanation for
      the excess absorption is that the X-rays are formed 
      within the wind and are strongly absorbed by the wind
      at lower energies.
   
\item The  presence of high-temperature plasma in supposedly
      single WN stars is not predicted by radiative wind shock models.
      Other mechanisms are thus required to explain the hotter
      plasma. Close companion models are not clearly justified
      given the lack of evidence for binarity. Magnetic wind 
      confinement models provide some promise, but the presence  
      of magnetic fields in WR stars has so far not been demonstrated.
      Further theoretical work is needed to develop models that can
      account for such hot plasma.      

\item Although the sample of WN stars observed in X-rays is
      small, there is a noticeable trend toward lower L$_{\rm X}$
      in the later WN7-9 subtypes. This is in the opposite sense
      of L$_{bol}$, which increases toward later WN7-9 subtypes.
      Thus, other factors besides  L$_{bol}$ determine X-ray
      luminosity in single WN stars. Wind properties appear to be
      important, and sensitive  X-ray observations of a broader sample of
      WN stars, particularly at low terminal wind speeds,
      are needed to determine if  L$_{\rm X}$ depends on
      wind parameters.

\end{enumerate}

%\placetable{tbl-1}
%\placetable{tbl-2}
%\placefigure{fig3}

\acknowledgments

This work was supported by {\em Chandra} award GO8-9008X issued
by the Chandra X-ray Observatory Center (CXC) and by
NASA/GSFC award NNX09AR25G. The CXC is
operated by the Smithsonian Astrophysical Observatory (SAO)
for, and on behalf of, the National Aeronautics Space
Administration under contract NAS8-03060.
We acknowledge use of {\em Chandra} archive data
for WR 24  obtained in a guaranteed time observation
(GTO ObdId  9482, PI G. Garmire). This observation was
part of a larger {\em Chandra} ACIS survey of the Carina
Nebula (Chandra Carina Complex Project; PI: L. Townsley).
This work was partially based on observations obtained with
XMM-Newton, an ESA science mission with instruments and
contributions directly funded by ESA member states
and the USA (NASA). SZ acknowledges financial support
from Bulgarian National Science Fund grant DO-02-85.

\clearpage

% That's the end of the main body of the paper.  Now we will have some
% back matter.
%
% Tables are supposed to be submitted one per page, following
% the main body of the text, so before each table we would have a
% \clearpage to force a page break at that point.  There should also
% be a \clearpage after the last table so that it gets forced onto
% its own page, too.
%
% 
% --------------------------------------------
% TABLE1
\begin{deluxetable}{llcccccc}
\tabletypesize{\scriptsize}
\tablewidth{0pc}
\tablecaption{WN Star Properties }
\tablehead{
\colhead{Name} &
\colhead{Sp. Type } &
\colhead{d} &
\colhead{A$_{\rm V}$ } &
\colhead{v$_{\infty}$} &
\colhead{$\dot{\rm M}$} &
\colhead{R$_{*}$ } &
\colhead{log L$_{bol}$} \\
\colhead{         } &
\colhead{         } &
\colhead{(kpc)} &
\colhead{(mag)} &
\colhead{(km/s)} &
\colhead{(10$^{-5}$ M$_{\odot}$/yr)} &
\colhead{(R$_{\odot}$)} &
\colhead{(L$_{\odot}$)}  \\
}
\startdata
\multicolumn{8}{c}{New Detections\tablenotemark{a}}   \nl
WR 2   & WN2    & 2.51  &  1.58   & 3200\tablenotemark{b}   & $\leq$2.0\tablenotemark{c} & 0.89 & 5.45 \nl
WR 18  & WN4    & 2.20  &  2.63    & 1800  & 2.5                    & 1.49 & 5.50  \nl
WR 79a & WN9ha  & 1.99  &  1.28                     & 955   & 2.4\tablenotemark{d}   & 33.1 & 5.78   \nl
WR 134 & WN6    & 1.74  &  1.52                     & 1820\tablenotemark{e}  & 4.0                    & 5.29 & 5.60  \nl
\multicolumn{8}{c}{Archive Data\tablenotemark{f}}   \nl
WR 16  & WN8h    & 2.37  & 1.80      & 650  & 5.0   & 19.9 & 6.15  \nl
WR 20b & WN6h    & $\geq$2.37  & 6.08   & ...  & ...      & ...  & 5.86\tablenotemark{g}  \nl
WR 24  & WN6ha   & 3.24  & 0.56      & 2160 & 4.0   & 19.9 & 6.35  \nl
WR 78  & WN7h    & 1.99  & 1.50      & 1385 & 5.0   & 16.7 & 6.20  \nl
WR 136 & WN6h    & 1.64\tablenotemark{h}  & 1.73       & 1600 & 3.2   & 3.3 & 5.40  \nl

\enddata
\tablenotetext{a}{
Spectral type, distance  and A$_{V}$
from vdH01 (A$_v$=1.11 A$_V$). For WR 2, WR 18, WR 134:  
v$_{\infty}$,  $\dot{\rm M}$,  R$_{*}$, L$_{bol}$  from Hamann et al. (2006),
unless otherwise noted.
For WR 79a: v$_{\infty}$, $\dot{\rm M}$,
R$_{*}$, L$_{bol}$ from  Crowther \& Bohannan (1997).
}
\tablenotetext{b}{Howarth \& Schmutz (1992) obtained
                  v$_{\infty}$ = 3200 km s$^{-1}$. Hamann et al. (2006)
                  give v$_{\infty}$ = 1800 km s$^{-1}$.  }
\tablenotetext{c}{Upper limit is based on a {\em VLA} non-detection with a
                  6 cm flux density S$_{6cm}$ $<$0.2 mJy (Abbott et al. 1986)
                  and v$_{\infty}$ = 3200 km s$^{-1}$.  } 
\tablenotetext{d}{Spectroscopic mass-loss rate (Crowther \& Bohannan 1997).}
\tablenotetext{e}{Ignace et al. (2001).}
\tablenotetext{f}{For archive objects, the spectral type, distance, and A$_{\rm V}$
                  are from vdH01 and other properties from Hamann et al. (2006),
                  unless otherwise noted. No mass-loss data specific for WR 20b were
                  found.}
\tablenotetext{g}{~L$_{bol}$ from Naz\'{e} et al. (2008).}
\tablenotetext{h}{From {\em Hipparcos} parallax (0.61 mas).}
\end{deluxetable}
\clearpage

% --------------------------------------------
% TABLE2
\begin{deluxetable}{lllll}
\tablewidth{0pc}
\tablecaption{WN Stars: New X-ray Observations }
\tablehead{
\colhead{Parameter} &
\colhead{WR 2} &
\colhead{WR 18} &
\colhead{WR 134} &
\colhead{WR 79a}  \\
}
\startdata
Date              & 2008 June 30  & 2008 Oct 29   & 2008 Feb 10 & 2009 Aug 30   \nl
ObsId             & 8943          & 8910          & 8909        & 0602020101      \nl
Start Time (TT)   & 06:17:17      & 16:59:38      & 10:23:29    & 01:44:03       \nl
Stop  Time (TT)   & 11:17:51      & 23:06:05      & 16:31:11    & 11:59:17       \nl
Exposure (s)      & 15,444        & 19,684        & 19,301      & 35,033 (pn), 36,617 (per MOS) \nl
Instrument        & ACIS-S        & ACIS-S        & ACIS-S\tablenotemark{a} & EPIC \nl
Frametime (s)     & 3.1           & 3.1           &  0.9        & 0.073 (pn), 2.6 (MOS)   \nl
\enddata
\tablenotetext{a}{WR 134 observed using 1/4 subarray to mitigate pileup.}
\end{deluxetable}
\clearpage
%
%
% ----------------------------------------
%% TABLE 3 
%
\begin{deluxetable}{llllllll}
\tabletypesize{\scriptsize}
\tablewidth{0pt} 
\tablecaption{WN Star X-ray Source Properties\tablenotemark{a}}
\tablehead{
	 \colhead{Name}                           &
           \colhead{R.A.}                         &
           \colhead{Decl.}                        &
           \colhead{Net Counts}                   &
           \colhead{Rate}                         & 
          \colhead{E$_{50}$}                      &
           \colhead{P$_{\rm const}$}              &
            \colhead{Identification(offset)}     \\
           \colhead{}                             &        
           \colhead{(J2000)}                      &
           \colhead{(J2000)}                      &
           \colhead{(cts)}                        &               
           \colhead{(cts/ksec)}                   & 
           \colhead{(keV)}                        &            
           \colhead{}                             & 
            \colhead{(arcsec)} 
                                  }
\startdata
\multicolumn{8}{c}{New Detections} \\
WR 2   & 01 05 23.05 & +60 25 18.9 & 247$\pm$16 & 16.0$\pm$1.04  & 1.04  & 0.68 (0.92)  & HST J010523.02+602518.9   (0.22)\\
WR 18  & 10 17 02.26 & -57 54 46.8 & 167$\pm$13 &  8.48$\pm$0.66 & 1.67  & 0.99 (0.91)  & HST J101702.28$-$575446.9 (0.19)\\
WR 79a & 16 54 58.52 & -41 09 02.9 & 805$\pm$29\tablenotemark{b} & 23.0$\pm$0.83\tablenotemark{b} & 1.35  & 0.63 (...) & HST J165458.50$-$410903.1 (0.30) \\
WR 134 & 20 10 14.19 & +36 10 35.1 & 785$\pm$28 & 40.7$\pm$1.45  & 1.90  & 0.95 (0.95)  & HST J201014.19+361035.1   (0.00)\\
\multicolumn{8}{c}{Chandra Archive Detections} \\
WR 20b\tablenotemark{c} & 10 24 18.39 & -57 48 29.9 &  643$\pm$26 & 13.0$\pm$0.53  & 3.04 & 0.24 (0.93)  & HST J102418.41$-$574829.7   (0.26)\\
WR 24\tablenotemark{d}  & 10 43 52.24 & -60 07 04.2 & 1419$\pm$39 & 25.1$\pm$0.69  & 1.66 & 0.67 (0.97)  & HST J104352.26$-$600704.0   (0.18)\\
WR 136\tablenotemark{e} & 20 12 06.56 & +38 21 18.2 &  238$\pm$17 & 2.56$\pm$0.18  & 1.96 & 0.99 (0.93)  & HST J201206.54$+$382117.8   (0.46)\\
\enddata
\tablenotetext{a}{
Notes:~{\em Chandra} X-ray data for WR 2, WR 18, WR 134, and WR 136 are from ACIS-S
and WR 20b, WR 24 are from ACIS-I.
 {\em XMM-Newton} data for WR 79a are from EPIC pn.  Parameters were determined using events in
 the 0.3 - 8 keV range.  Tabulated quantities are: 
source name, J2000.0 X-ray position (R.A., Decl.), net counts and 
net counts error (from 3$\sigma$ {\em wavdetect} ellipses for {\em Chandra} sources and from a 90\% encircled
energy circular source 
region of radius r = 45$''$ for the {\em XMM} source WR79a, accumulated in the  exposure given in Table 2, 
rounded to the nearest integer,
background subtracted and PSF-corrected), count rate (Rate) obtained by dividing net counts by the exposure 
time in Table 2, median photon  energy  E$_{50}$, probability of constant count-rate (P$_{\rm const}$) 
from the KS test followed in parentheses by the corresponding probability from the Gegory-Loredo method
for {\em Chandra} sources as determined from the CIAO tool {\em glvary}, and HST GSC v2.3.2 optical counterpart identification
within a 2$''$ search radius. The offset (in arcsecs) between the X-ray 
and optical counterpart position is given in parentheses. 
}
\tablenotetext{b}{The corresponding EPIC MOS values in the 0.3 - 8 keV range using a r = 45$''$ source extraction region are: 
                  MOS1  286 $\pm$ 17 net counts, 7.81 $\pm$ 0.46 c/ksec;
                  MOS2  281 $\pm$ 17 net counts, 7.67 $\pm$ 0.46 c/ksec.}
\tablenotetext{c}{{\em Chandra} ObsID 6411, 49,377 s exposure on 28 Sept. 2006 ;WR 20b positioned $\approx$4.1$'$ off-axis.}
\tablenotetext{d}{{\em Chandra} ObsID 9482, 56,519 s exposure on 18 Aug. 2008; WR 24 positioned $\approx$7.4$'$ off-axis.}
\tablenotetext{e}{{\em Chandra} ObsID 3763, 92,771 s exposure on 19 Feb. 2003; WR 136 positioned $\approx$8.2$'$ off-axis.}
\end{deluxetable}
\clearpage
% --------------------------------------------
% TABLE4.TEX 
\begin{deluxetable}{lllll}
\tabletypesize{\scriptsize}
%\tablewidth{33pc}
\tablewidth{0pc}
\tablecaption{X-ray Spectral Fits of WN Stars (New Detections)  
   \label{tbl-1}}
\tablehead{
\colhead{Parameter}      &
\colhead{ }        &
\colhead{Object}        &
\colhead{ }        &
\colhead{  }
}
\startdata
Star                                & WR 2                & WR 18               & WR 134             & WR79a           \nl
Model                               & 2T vapec            & 2T vapec           & 2T vapec            & 2T vapec        \nl
Abundances\tablenotemark{a}         & WN\tablenotemark{b} & WN\tablenotemark{c} & WN\tablenotemark{d}& WN\tablenotemark{e}\nl
N$_{\rm H}$ (10$^{22}$ cm$^{-2}$)   & 0.44 [0.08 - 0.99]  & 1.37 [0.72 - 1.86]  & 1.19 [0.83 - 1.45] & 0.75 [0.64 - 0.83] \nl
kT$_{1}$ (keV)                      & 0.32 [0.24 - 0.45]  & 0.65 [0.43 - 0.82]  & 0.62 [0.53 - 0.71] & 0.55 [0.46 - 0.61] \nl
norm$_{1}$ (10$^{-6}$)              & 1.73 [0.99 - 70.0]  & 2.24 [0.45 - 7.85]  & 1.47 [0.34 - 3.70] & 2.04 [1.39 - 3.16] \nl 
kT$_{2}$ (keV)                      & $\geq$2.6\tablenotemark{f}  & 3.51 [1.86 - ...]   & 8.57 [5.74 - 17.2]   & 2.66 [1.66 - 5.60] \nl
norm$_{2}$  (10$^{-6}$)             & 0.56 [0.14 - 0.89]  & 0.85 [0.33 - 1.48]  & 3.75 [1.66 - 5.04] & 0.47 [0.27 - 0.73]  \nl
$\chi^2$/dof                        & 17.2/14             & 4.0/7                    & 57.6/56       &  46.4/48       \nl
$\chi^2_{red}$                      & 1.23\tablenotemark{g}  & 0.57                  & 1.03          &  0.97          \nl
F$_{\rm X}$ (10$^{-13}$ ergs cm$^{-2}$ s$^{-1}$)          & 0.86 (4.56) & 0.58 (4.87) & 4.36 (12.5)  &  0.64 (2.94)   \nl
F$_{\rm X,1}$ (10$^{-13}$ ergs cm$^{-2}$ s$^{-1}$)        & 0.37 (3.96) & 0.22 (4.16) & 0.68 (7.20)  &  0.41 (2.53)   \nl
log L$_{\rm X}$ (ergs s$^{-1}$)     & 32.54               & 32.45                     & 32.66        &  32.14         \nl
log [L$_{\rm X}$/L$_{bol}$]         & $-$6.49             & $-$6.63                   & $-$6.52      &  $-$7.22       \nl
log [L$_{\rm X}$/L$_{wind}$]        & $\geq$$-$5.27          & $-$4.96   & $-$4.96      &  $-$4.70       \nl
N$_{\rm H}$/N$_{\rm H,vis}$         & 1.3 [0.2 - 2.8]       & 2.3 [1.2 - 3.2]           & 3.5 [2.5 - 4.3]  & 2.6 [2.2 - 2.9] \nl    
\enddata
\tablecomments{
Based on  XSPEC (vers. 12.4.0) fits of the background-subtracted  spectra binned 
to a minimum of 10 counts per bin using the optically thin thermal plasma
$vapec$ model. All models included the XSPEC $wabs$ photoelectric absorption component,
based on Morrison \& McCammon (1983) cross-sections and Anders \& Ebihara (1982)
relative abundances. The tabulated parameters
are equivalent neutral H absorption column density (N$_{\rm H}$), the product of 
Boltzmann's constant time plasma temperature  (kT),
and XSPEC component normalization (norm). 
%%Gaussian line
%%centroid energy (E$_{line}$), line width ($\sigma_{line}$ = FWHM/2.35).
Square brackets enclose 90\% confidence intervals and an ellipsis means that 
the algorithm used to compute confidence intervals did not converge.
The total X-ray flux (F$_{\rm X}$) and flux of the cool component
(F$_{\rm X,1}$) are the absorbed values in the 0.3 - 8 keV range, followed in 
parentheses by  unabsorbed values. 
%%The continuum-subtracted Fe  line flux
%%(F$_{\rm X,line}$) is measured in the 6.5 - 6.84 keV range.
The unabsorbed luminosity L$_{\rm X}$ (0.3 - 8 keV)  assumes the
distances given in Table 1.  L$_{bol}$ is from Table 1.
L$_{wind}$ = (1/2)$\dot{\rm M}$v$_{\infty}^2$ (Table 1).
The ratio N$_{\rm H}$/N$_{\rm H,vis}$ of X-ray to optically-derived
absorption is computed using 
N$_{\rm H,vis}$ = 2.22 $\times$ 10$^{21}$A$_{\rm V}$ cm$^{-2}$
(Gorenstein 1975), where A$_{\rm V}$ is from Table 1.
}
\tablenotetext{a}{Abundances were held fixed at the generic WN 
values given in Table 1 of VCW86, except for specific elements
whose abundances were allowed to vary as noted below. The
generic WN abundances reflect H depletion and N enrichment and
are by number:
He/H = 14.9, C/H = 1.90E-03, N/H = 9.36E-02,
O/H = 4.35E-03, Ne/H = 9.78E-03, Mg/H = 3.26E-03, 
Si/H = 3.22E-03, P/H = 1.57E-05, S/H = 7.60E-04,
Fe/H = 1.90E-03. All other elements were held fixed
at solar abundances (Anders \& Grevesse 1989). }

\tablenotetext{b}{The abundances of N, Mg, and Fe were varied 
to improve the fit and
converged to values  N = 5.3, Mg = 2.1, and Fe = 1.3 times  
the  number ratios given in VCW86. The S abundance is poorly
constrained and was held fixed at S = 3.0 times the number ratio 
given in VCW86.}

\tablenotetext{c}{The abundances of Mg and Fe were varied 
to improve the fit and converged to values Mg = 0.3, and 
Fe = 2.1 times the number ratios given in VCW86. }

\tablenotetext{d}{The abundances of Mg, Si, S, and Fe were varied
to improve the fit and converged to values  
Mg = 3.3, Si = 6.0, S = 18.6, and Fe = 5.6 
times the number ratios given in VCW86.}

\tablenotetext{e}{The abundances of  Si, S, and Ne were varied
to improve the fit and converged to values  
Si = 2.7, S = 8.4, Ne = 0.46 times the number ratios given in 
VCW86. A solar abundance 2T $vapec$ fit underestimates the 
Si XIII line flux, but allowing the Si abundance to vary
gives an acceptable fit with Si = 3.4 $\times$ solar.
The temperatures inferred from the solar abundance fit
are nearly identical to those with WN abundances, but
solar abundances converge to an N$_{\rm H}$ value that
is 20\% larger and give a slightly larger 
log L$_{\rm X}$ = 32.3 ergs s$^{-1}$.
}

\tablenotetext{f}{~Lower bound is 90\% confidence. The value
of kT$_{2}$  is sensitive to the  S abundance and
not tightly-constrained.}

\tablenotetext{g}{Some further reduction in $\chi^2_{red}$ can be obtained
 by allowing the S abundance to increase to very large values.}

\end{deluxetable}
\clearpage

% TABLE5.TEX 
\begin{deluxetable}{llll}
\tabletypesize{\scriptsize}
%\tablewidth{33pc}
\tablewidth{0pc}
\tablecaption{X-ray Spectral Fits of WN6 Stars ({\em Chandra} Archive Data)  
   \label{tbl-1}}
\tablehead{
\colhead{Parameter}      &
\colhead{ }        &
\colhead{Object}        &
\colhead{  }
}
\startdata
Star                                & WR 20b              & WR 24               & WR 136              \nl
Model                               & 2T vapec            & 2T vapec            & 2T vapec            \nl
Abundances\tablenotemark{a}         & WN\tablenotemark{b} & WN\tablenotemark{c} & WN\tablenotemark{d} \nl
N$_{\rm H,1}$ (10$^{22}$ cm$^{-2}$) & 3.92 [2.96 - 7.45]  & 0.82 [0.43 - 1.06]  & 0.49 [0.36 - 0.72]  \nl
kT$_{1}$ (keV)                      & 0.44 [0.28 - 0.75]  & 0.70 [0.64 - 0.83]  & 0.56 [0.33 - 0.73]  \nl
norm$_{1}$ (10$^{-6}$)              & 54.6 [24.9 - 76.0]  & 5.61 [1.63 - 8.88]  & 20.3 [10.2 - 103.]  \nl 
N$_{\rm H,2}$ (10$^{22}$ cm$^{-2}$) & ...                 & ...                 & 3.98 [1.63 - 7.41]  \nl
kT$_{2}$ (keV)                      & 5.43 [3.16 - 9.09]  & 3.33 [2.48 - 4.83]  & 2.64 [1.53 - 18.0]    \nl
norm$_{2}$  (10$^{-6}$)             & 5.56 [4.46 - 8.23]  & 3.22 [2.32 - 4.36]  & 87.0 [59.8 - 205.]   \nl
$\chi^2$/dof                        & 55.6/50             & 110.2/99            & 49.6/46         \nl
$\chi^2_{red}$                      & 1.11\tablenotemark{e}  & 1.11                & 1.08             \nl
F$_{\rm X}$ (10$^{-13}$ ergs cm$^{-2}$ s$^{-1}$)          & 2.27 (60.6) & 2.51 (7.95) & 0.29 (1.02)     \nl
F$_{\rm X,1}$ (10$^{-13}$ ergs cm$^{-2}$ s$^{-1}$)        & 0.32 (56.0) & 1.10 (5.45) & 0.07 (0.27)     \nl
log L$_{\rm X}$ (ergs s$^{-1}$)     & $\geq$33.57            & 33.0        & 31.51           \nl
N$_{\rm H,1}$/N$_{\rm H,vis}$         & 2.9 [2.2 - 5.5]     & 6.6 [3.5 - 8.5]           & 1.3 [0.9 - 1.9]   \nl 
N$_{\rm H,2}$/N$_{\rm H,vis}$         & ...                 & ...                       & 10.4 [4.2 - 19.3]   \nl
\enddata
\tablecomments{
Based on  XSPEC (vers. 12.4.0) fits of the background-subtracted  spectra binned 
to a minimum of 10 counts per bin using the optically thin thermal plasma
$vapec$ model. The tabulated parameters are the same as in Table 4.
Square brackets enclose 90\% confidence intervals.
The total X-ray flux (F$_{\rm X}$) and flux of the cool component
(F$_{\rm X,1}$) are the absorbed values in the 0.3 - 8 keV range, followed in 
parentheses by  unabsorbed values. 
The unabsorbed luminosity L$_{\rm X}$ (0.3 - 8 keV)  assumes the
distances given in Table 1.  
}
\tablenotetext{a}{Abundances were held fixed at the generic WN 
values given in Table 1 of VCW86, except for specific elements
whose abundances were allowed to vary as noted below. The
generic WN abundances are given in Table 4.}

\tablenotetext{b}{The abundances of Mg, Si, and S were varied 
to improve the fit and
converged to values Mg = 0.6, Si = 0.2, and S = 3.0 times  
the  number ratios given in VCW86.}

\tablenotetext{c}{The abundances of Mg, Si, S,  and Fe were varied 
to improve the fit and converged to values Mg = 0.9, Si = 1.1, 
S = 5.2, and  Fe = 0.6 times the number ratios given in VCW86. }

\tablenotetext{d}{The abundances of Si and S were varied
to improve the fit and converged to values  
Si = 2.6 and  S = 5.4 
times the number ratios given in VCW86.}

\tablenotetext{e}{A slightly better fit can be obtained using the
 plane-parallel shock model $vpshock$. This fit converges to
 N$_{\rm H}$ = 4.67 [3.67 - 6.11] $\times$ 10$^{22}$ cm$^{-2}$,
 kT$_{shock}$ = 3.86 [2.97 - 6.53] keV, Mg = 1.7, Si = 0.3,
S = 1.6 times the number ratios given in VCW86,
$\chi^2$/dof = 45.5/51, $\chi^2_{red}$ = 0.89,
F$_{\rm X}$(0.3 - 8 keV) = 2.09 (97.1) $\times$ 10$^{-13}$
ergs cm$^{-2}$ s$^{-1}$, log L$_{\rm X}$ = 33.78 ergs s$^{-1}$
(d = 2.27 kpc).}

\end{deluxetable}
\clearpage

\clearpage

%%*** Uncomment the following line to skip the figures ***
%%\end{document}
%%*******************************************************

\begin{figure}
\figurenum{1}
\includegraphics*[width=7.0cm,angle=-90]{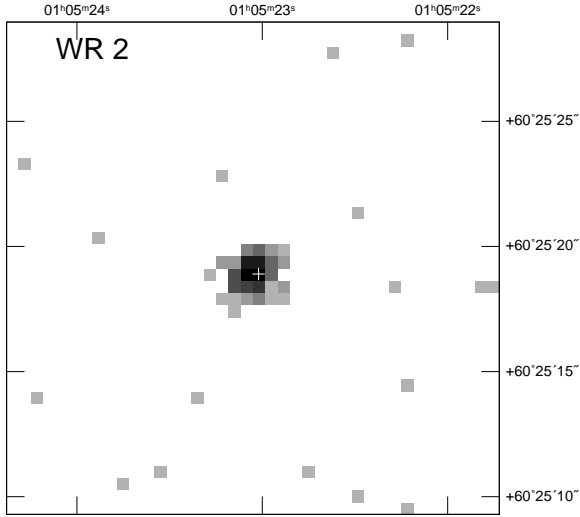}
\caption{{\em Chandra} ACIS-S image of WR 2 in the 0.3 - 8 keV
 energy range. The pixel size is 0.$''$492. The image is displayed 
 on a  log intensity scale and the coordinates are J2000.0. The
 cross marks the {\em HST} GSC position of WR 2. }
\end{figure}

%%\clearpage

\begin{figure}
\figurenum{2}
\includegraphics*[width=7.0cm,angle=-90]{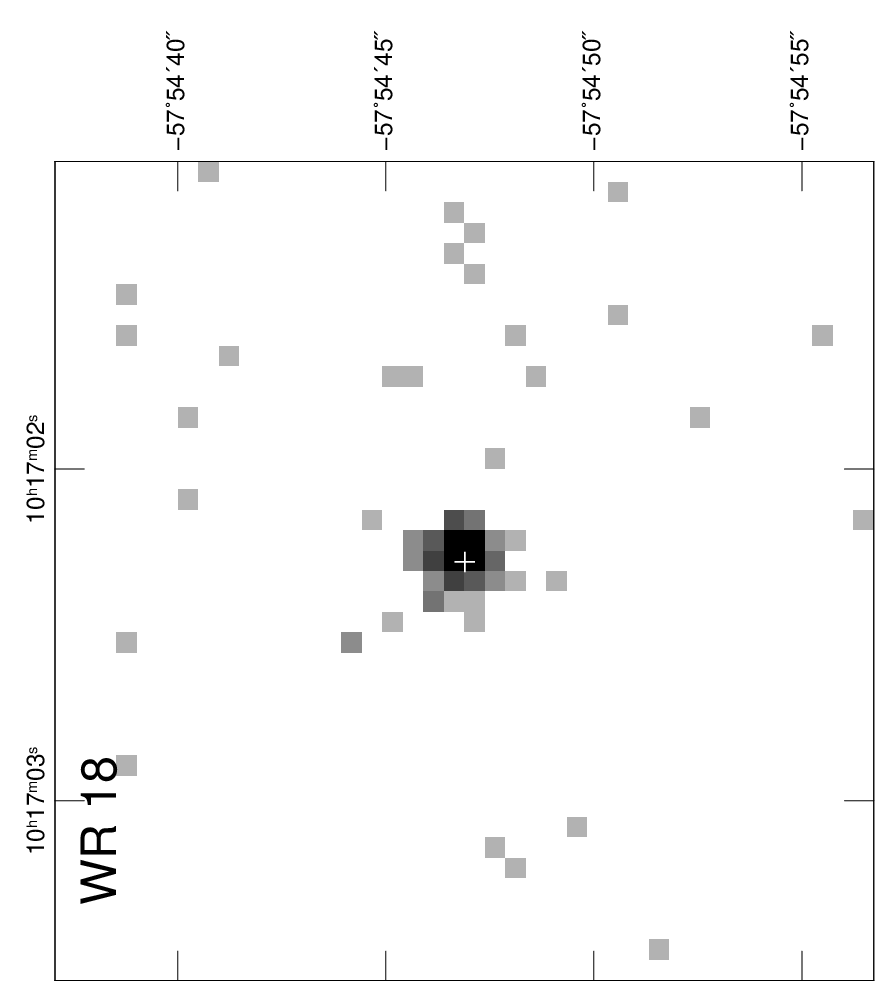}
\caption{{\em Chandra} ACIS-S image of WR 18 in the 0.3 - 8 keV 
energy range. Annotation and 
image properties are the same as in Figure 1.}
\end{figure}

\clearpage

\begin{figure}
\figurenum{3}
\includegraphics*[width=7.0cm,angle=-90]{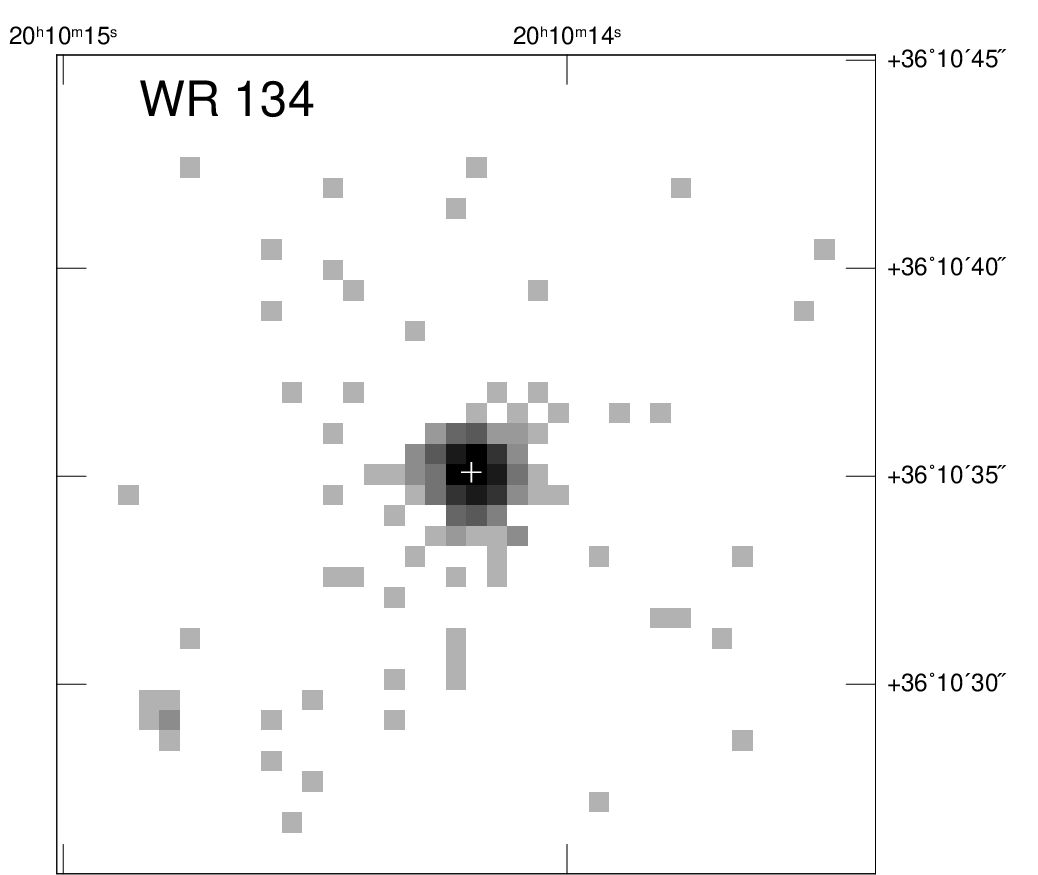}
\caption{{\em Chandra} ACIS-S image of WR 134 in the 0.3 - 8 keV
 energy range. Annotation and 
image properties are the same as in Figure 1.}
\end{figure}

\begin{figure}
\figurenum{4}
\includegraphics*[width=7.0cm,angle=-90]{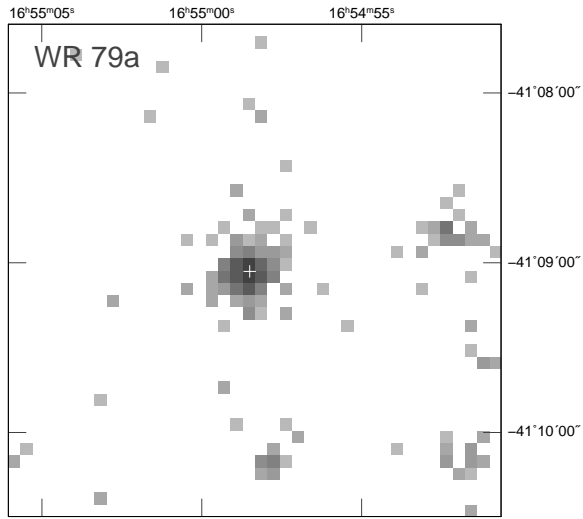}
\caption{{\em XMM-Newton} EPIC pn image of WR 79a in the 0.3 - 8 keV
 energy range. The pixel size is 4.$''$3. The image is displayed 
 on a  log intensity scale and the coordinates are J2000.0. The
cross marks the {\em HST} GSC position of WR 79a.}
\end{figure}

\clearpage

\begin{figure}
\figurenum{5}
\includegraphics*[width=12.0cm,angle=-90]{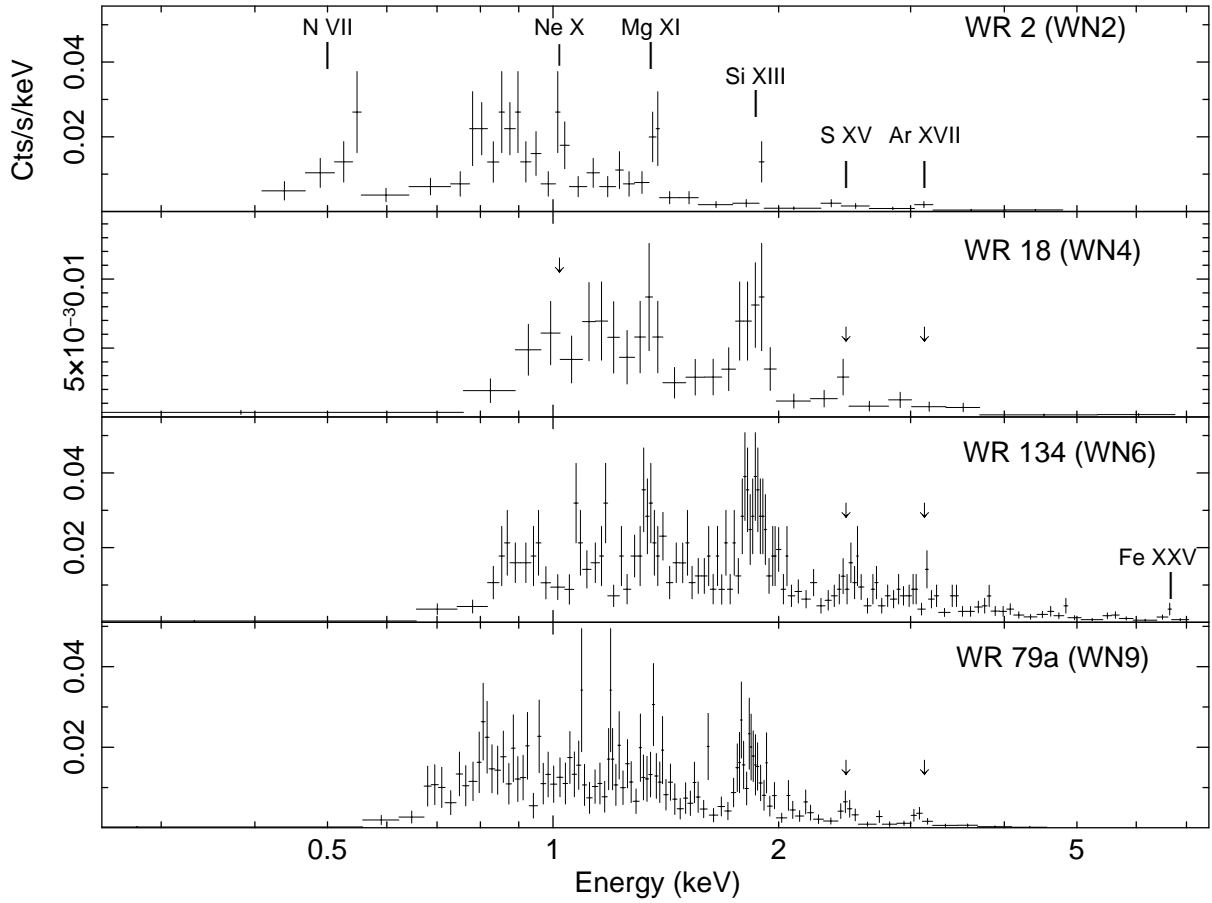}
\caption{{\em Chandra} ACIS-S  spectra of 
WR 2, WR 18, and WR 134 and {\em XMM-Newton} EPIC pn spectrum of
WR 79a.  The spectra are background-subtracted and rebinned to a minimum of
5 counts per bin for display. Prominent emission lines are marked.}
\end{figure}

\clearpage

\begin{figure}
\figurenum{6}
\includegraphics*[width=12.0cm,angle=-90]{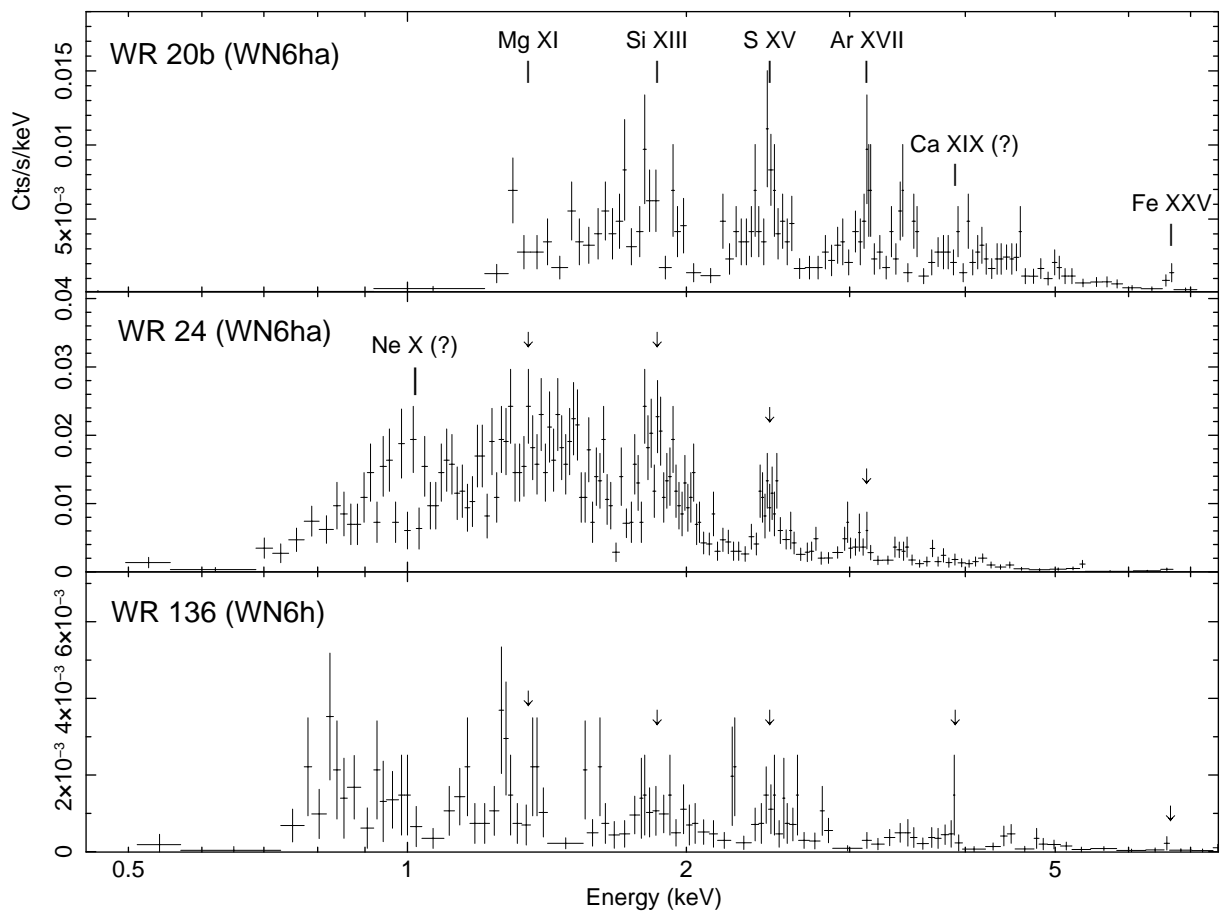}
\caption{Background-subtracted {\em Chandra} ACIS spectra of 
WR 20b, WR 24, and WR 136 from  archive data.
The spectra of WR 20b and WR 24 are rebinned to a minimum of
5 counts per bin, and the fainter source WR 136 is rebinned 
to a minimum of 2 counts per bin. Prominent emission lines are 
marked. The Ca XIX line in WR 20b and WR 136 is faint and 
is classified as a possible detection.}
\end{figure}

\clearpage

\begin{figure}
\figurenum{7}
\includegraphics*[width=10.0cm,angle=-90]{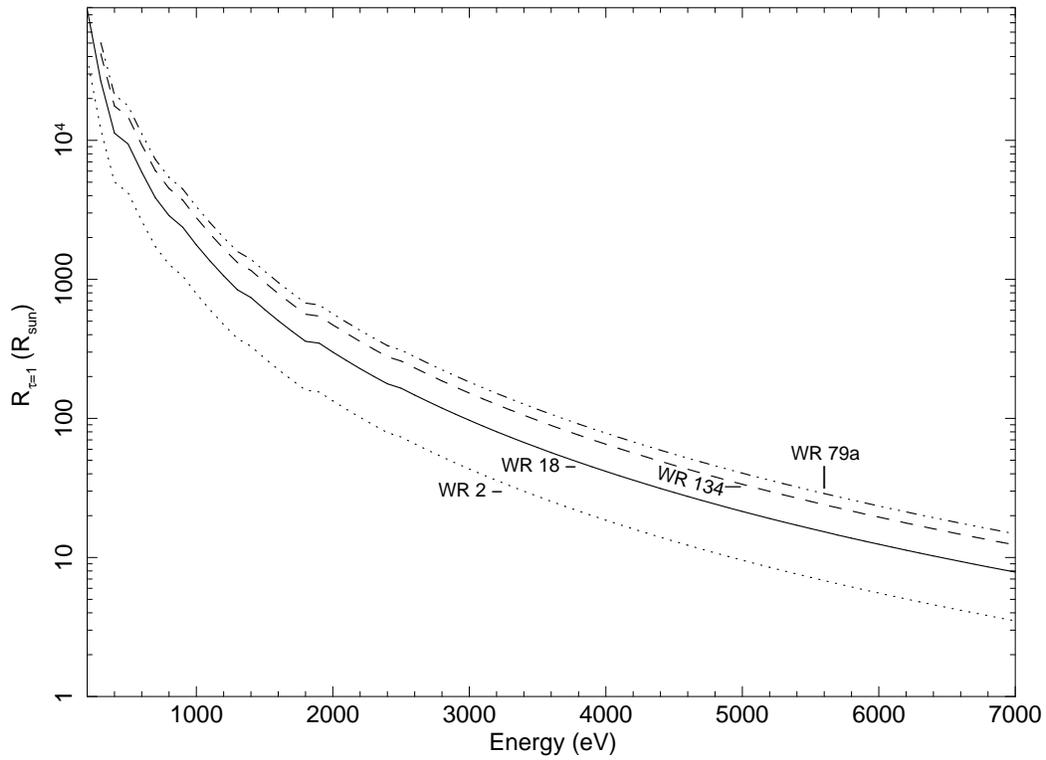}
\caption{Radius of optical depth unity as a function of photon
         energy for the detected WN stars. The radii are calculated
         for a spherical constant-velocity wind
         using the mass-loss parameters in Table 1, generic WN
         abundances (VCW86), and BM92 cross-sections.}
\end{figure}

\clearpage

\begin{figure}
\figurenum{8}
\includegraphics*[width=10.0cm,angle=-90]{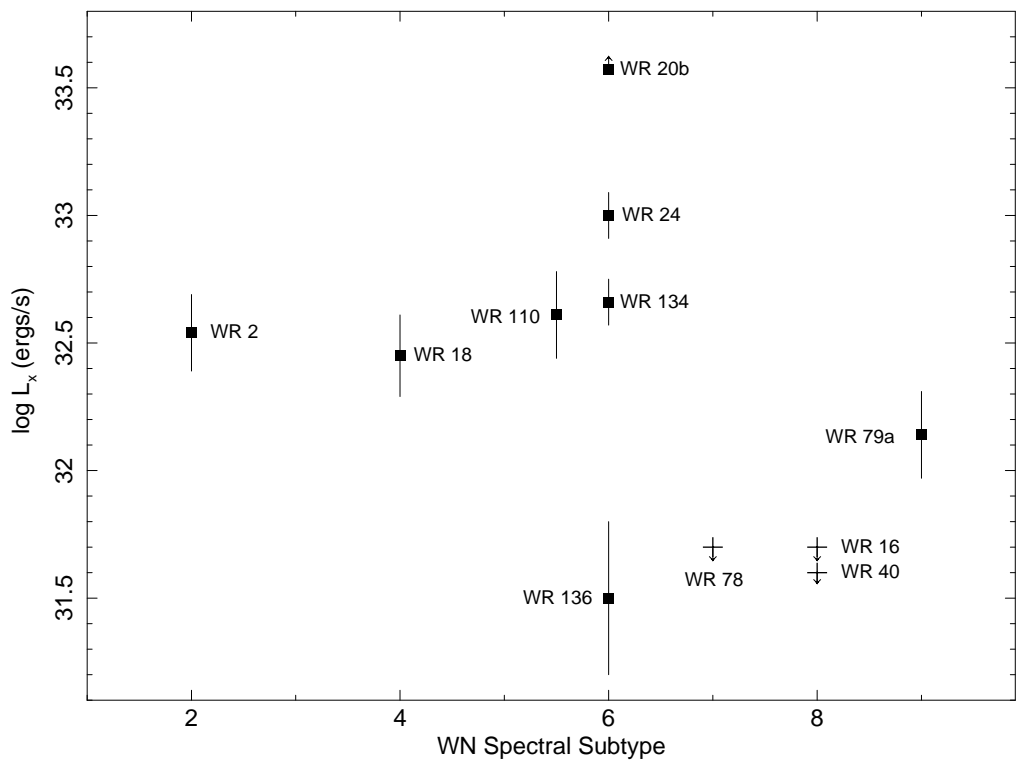}
\caption{X-ray luminosity (Tables 4 and 5) versus spectral subtype for
putatively single WN stars. The X-ray data for
WR 110 (WN5-6) are from {\em XMM-Newton} (Skinner et al. 2002a).
The X-ray upper limit for WR 40 (WN8) is from 
{\em XMM-Newton} (Gosset et al. 2005).
The  3$\sigma$ upper limits for WR 16 (WN8h) and 
WR 78 (WN7h) are based on analysis of archival {\em ROSAT} PSPC 
images (Sec. 3.3). Error bars on L$_{\rm X}$ are internal
only and do not include possible systematic effects such as distance
uncertainties.
}
\end{figure}

\clearpage

\begin{figure}
\figurenum{9}
\includegraphics*[width=10.0cm,angle=-90]{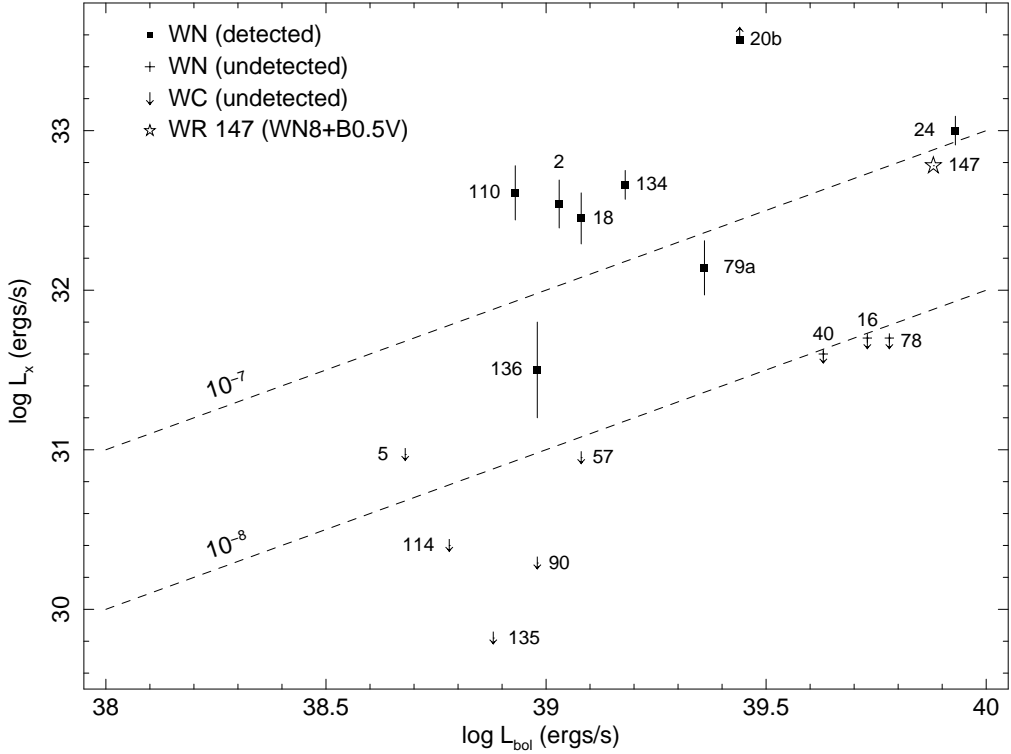}
\caption{X-ray versus bolometric  luminosity for suspected
single WN and WC stars. The WN $+$ OB binary system WR 147 is also
shown for comparison. L$_{bol}$ values are from Hamann et al. (2006)
except for WR 79a (Crowther \& Bohannan 1997) and 
WR 20b (Naz\'{e} et al. 2008). L$_{\rm X}$ for detections are
from Tables 4 and 5 except for WR 110 (Skinner et al. 2002a)
and WR 147 (Skinner et al. 2007). The WN4 star WR 6 (= EZ CMa) is
an X-ray source (Skinner et al. 2002b), but is not shown because 
of an uncertain distance d = 0.58 - 1.8 kpc and suspected binarity. 
If an intermediate distance d = 1.2 kpc is assumed, it would lie 
at (L$_{bol}$, L$_{\rm X}$) $\approx$ (39.18, 32.7), which is nearly  
coincident with WR 134.  Plotted error bars on L$_{\rm X}$
are internal only and do not account for possible systematic
effects such as distance uncertainties.
X-ray upper limits for WC stars
are from Skinner et al. (2006, 2009), except for
WR 114 (Oskinova et al. 2003). The X-ray upper limit for
WR 40 is from Gosset et al. (2005). The dashed lines
show slopes for reference and are not regression fits.
} 
\end{figure}

\clearpage

\begin{figure}
\figurenum{10}
\includegraphics*[width=10.0cm,angle=-90]{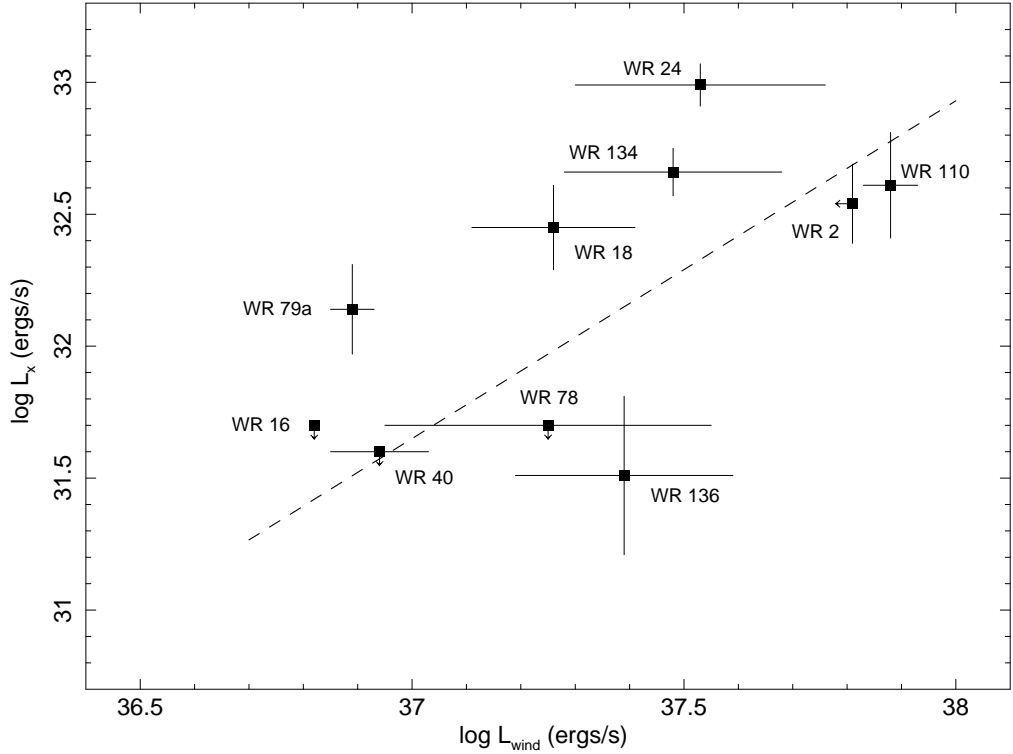}
\caption{X-ray versus wind luminosity (L$_{wind}$ =
$\frac{1}{2}$$\dot{\rm M}$v$_{\infty}^2$) for the   WN stars
in this study.  
Also included is the WN5-6 star WR 110 using data from 
Skinner et al. (2002a). WR 20b is excluded for lack of reliable
mass-loss data and an uncertain distance. 
The WN4 star WR 6 (= EZ CMa) is
an X-ray source (Skinner et al. 2002b), but is not shown because 
of an uncertain distance d = 0.58 - 1.8 kpc and suspected binarity. 
If an intermediate distance d = 1.2 kpc is assumed, along with
mass-loss estimates from Schmutz (1997), it would lie at 
at (L$_{wind}$, L$_{\rm X}$) $\approx$ (37.4, 32.7), which is nearly  
coincident with WR 134. 
The X-ray upper limits for WR 16 and WR 78 are 
from {\em ROSAT} (Sec. 3.3). The X-ray upper limit for 
WR 40 is from Gosset et al. (2005).  Error bars on L$_{\rm X}$
are internal only and do not account for possible systematic
effects such as distance uncertainties.
The plotted values of L$_{wind}$ are log averages based on
calculations using published $\dot{M}$ and v$_{\infty}$ values.
The error bars on L$_{wind}$ reflect the spread in published
$\dot{M}$ and v$_{\infty}$ for each star.  
The dashed  line has slope $m$ = $+$1.28 $\pm$ 0.51 and is a
regression fit to the data, taking both X-ray detections and upper limits
into account. {\em Notes (and mass-loss references in parentheses):}
WR 2:~ The upper limit on  L$_{wind}$  assumes 
log $\dot{\rm M}$ $\leq$ $-$4.7 (1) computed using 
v$_{\infty}$ = 3200 km s$^{-1}$ (2); 
WR 16:~(4); 
WR 18:~(3,4);
WR 24:~(3,4); 
WR 40:~(4,9)
WR 78:~(3,4);
WR 79a:~(3,5);
WR 134:~(1,2,4,6);
WR 136:~(1,4,6,7,8).~{\em References}: 
(1) Abbott et al. 1986;
(2) Howarth \& Schmutz 1992;
(3) Crowther 2007;
(4) Hamann et al. 2006;
(5) Crowther \& Bohannan 1997;
(6) Ignace et al. 2001;
(7) Ignace et al. 2003b;
(8) St.-Louis et al. 2009;
(9) Gosset et al. 2005.
}
\end{figure}

\end{document}